\documentclass[twocolumn,trackchanges]{aastex631}

\newcommand{\kms}{\hbox{km\,s$^{-1}$}} 
\newcommand{\logg}{\hbox{log\,$\it g$}}
\newcommand{\feh}{\hbox{$\rm [Fe/H]$}}
\newcommand{\teff}{\hbox{$T_{\rm eff}$}}
\newcommand{\vsini}{\hbox{$v$ \rm sin $i$}}
\usepackage{lmodern}
\usepackage[T1]{fontenc}
\usepackage{textcomp}
\usepackage{graphicx}
\usepackage{subfigure}
\usepackage{orcidlink}

\shorttitle{\vsini~ for LAMOST DR9}
\shortauthors{Zuo et al.}

\begin{document}
\title{Projected rotational velocities for LAMOST stars with effective temperature lower than 9000 K}

\correspondingauthor{ A-Li Luo}
\email{ lal@nao.cas.cn}

\author{Fang Zuo$^{\orcidlink{0000-0002-9081-8951}}$}
\affiliation{CAS Key Laboratory of Optical Astronomy, National Astronomical Observatories, Beijing 100101, China}
\affiliation{School of Astronomy and Space Science, University of Chinese Academy of Sciences, Beijing 101408, China}
\author{A-Li Luo$^{\orcidlink{0000-0001-7865-2648}}$}
\affiliation{CAS Key Laboratory of Optical Astronomy, National Astronomical Observatories, Beijing 100101, China}
\affiliation{School of Astronomy and Space Science, University of Chinese Academy of Sciences, Beijing 101408, China}
\affiliation{University of Chinese Academy of Sciences, Nanjing, Jiangsu, 211135, China}
\author{Bing Du$^{\orcidlink{0000-0001-6820-6441}}$}
\affiliation{CAS Key Laboratory of Optical Astronomy, National Astronomical Observatories, Beijing 100101, China}
\author{Yinbi Li$^{\orcidlink{0000-0001-7607-2666}}$}
\affiliation{CAS Key Laboratory of Optical Astronomy, National Astronomical Observatories, Beijing 100101, China}
\author{Hugh R. A. Jones$^{\orcidlink{0000-0003-0433-3665}}$}
\affiliation{School of Physics, Astronomy and Mathematics, University of Hertfordshire, College Lane, Hatfield AL10 9AB, UK}
\author{Yi-han Song} 
\affiliation{CAS Key Laboratory of Optical Astronomy, National Astronomical Observatories, Beijing 100101, China}
\author{Xiao Kong$^{\orcidlink{0000-0001-8011-8401}}$}
\affiliation{CAS Key Laboratory of Optical Astronomy, National Astronomical Observatories, Beijing 100101, China}
\author{Yan-xin Guo$^{\orcidlink{0000-0002-7640-5368}}$}
\affiliation{CAS Key Laboratory of Optical Astronomy, National Astronomical Observatories, Beijing 100101, China}

\begin{abstract}
%

In Data Release 9 of LAMOST, we present measurements of  \vsini~for a total of 121,698 stars measured using the Medium Resolution Spectrograph (MRS) and 80,108 stars using the Low Resolution Spectrograph (LRS). These values were obtained through a $\chi^2$ minimisation process,  comparing LAMOST spectra with  corresponding grids of synthetically broadened spectra. Due to the resolution and the spectral range of LAMOST,  \vsini~measurements are limited to stars with effective temperature (\teff) ranging from 5000 K to 8500 K for MRS and 7000 K to 9000 K for LRS. The detectable \vsini~ for MRS is set between 27 \kms~and 350 \kms~, and for LRS between  110 \kms~and 350 \kms, This limitation is because the convolved reference spectra become less informative beyond 350 \kms. The intrinsic precisions of \vsini~,  determined from multi-epoch observations, is approximately $\sim$ 4.0 \kms~for MRS and $\sim$ 10.0 \kms~for LRS at signal-to-noise ratio (S/N) greater than  50. Our \vsini~values show consistence with those from APOGEE17, displaying a scatter of 8.79 \kms. They are also in agreement with measurements from the Gaia DR3 and SUN catalogs. An observed trend in LAMOST MRS data is the decrease in  \vsini~ with dropping \teff, particularly transiting around 7000~K for dwarfs and 6500~K for giants,  primarily observed in stars with near-solar abundances.
\end{abstract}

\section{Introduction \label{sec:intro}}

Measuring stellar rotation is of long-standing interest in stellar astrophysics. It is  a key parameter for detecting the evolution of stellar angular momentum, which is advocated as the mechanism responsible to explain mixing and dilution \citep{2021ApJ...911..138N, 2021ApJS..257...22S,2023AJ....165..182K}. Thus, knowledge of stellar rotation velocity may provide us a better understanding of stellar interior structure. In addition, rotation period is an important parameter for Gyrochronology, age-dating using a star's rotation period and mass, that is mostly important for cool stars that evolve too slowly for isochrone dating to work \citet{2023arXiv231014990L}.

Several methods have been used to determine projected rotational velocity (\vsini) from spectroscopy. It can be obtained from the first zero frequency of the Fourier transform (FT) of the isolated spectral line profile \citep{1976PASP...88..809S,2002A&A...393..897R,2014ApJ...797...29L,2020PASJ...72...10T}, or the cross-correlation function (CCF) technique \citep{2001A&A...375..851M}. However, the two methods were not widely applied to large sky survey projects, because the FT technique needs high resolution and high signal-to-noise ratios (S/Ns) spectra, and the CCF technique requires a calibrator with \vsini~accurately determined by other techniques. One widely used method was proposed in \cite{1949ApJ...110..498S}, which use a rotational broadening function to generate a theoretical grid with different rotations, and determined \vsini~of 125 brighter O, B2-B5 and B2e-B5e stars by comparing the observed line contours of He I 4026 \AA~to the theoretical line contours.  \citet{1954ApJ...119..146S, 1955ApJ...121..653S} conducted subsequent measurement for several hundred stars ranging from B8 to G0, adhering to his previous method. In 1975, he further extended his research by obtaining \vsini~measurements for 217 stars classified as O9-F9, which serve as a calibration standard for rotational velocities \citep{1975ApJS...29..137S}. This method has also been applied to determining the rotational velocities of cool stars with high-resolution spectra. \citep{1998A&A...331..581D} derived the \vsini~values for 118 field M dwarfs, and \cite{2008ApJ...684.1390R} determined the \vsini~values for 45 L dwarfs. 

As the \vsini~values for an increasing number of stars were calculated, several interesting phenomena emerged. \cite{1949ApJ...110..498S} discovered that Be stars have rotation velocities larger than those of typical B-type stars. \cite{1982PASP...94..271F} found that the Ae stars are faster rotators compared to typical A-type stars, and \cite{1979A&A....74...38B} found that chemically peculiar (CP) metallic-line (Am) stars are predominantly slow rotators. For dwarf stars, \cite{1967ApJ...150..551K} discovered the ‘Kraft break’ around F0, and stars hotter than F0 have faster rotational velocities. The break was also discovered by the later spectroscopic studies \citep{2019ApJ...876..113S, 2020MNRAS.492.2177K}. For giant stars, \cite{1989ApJ...347.1021G} found the rotational break occurs near G0. The reason for this observable variations may be that hot stars possess extremely thin convective envelopes, which prevents them from generating magnetic winds and causing angular momentum loss, while cold stars have thick convective envelopes thus their slow rotation velocity is considered a consequence of magnetic braking \citep{1962AnAp...25...18S,1988ApJ...333..236K}. It is known that stellar rotation is a complex physical process and a large sample of rotational velocity catalog from large sky surveys allow us to better understand the multifaceted nature of stellar rotation.



With the development of large sky surveys, data reduction pipelines were being developed to automatically measure stellar parameters. The Apache Point Observatory Galactic Evolution Experiment (APOGEE) developed the APOGEE Stellar Parameter and Chemical Abundances Pipeline (ASPCAP) for the automated analysis of high-resolution spectra (R $\sim$ 22,500 across 15,100 -- 15,799 \AA, 15,867 -- 16,424 \AA, and 16,484 -- 17,000 \AA) of the stars across the Milky Way \citep{2016AJ....151..144G}. ASPCAP determines atmospheric parameters (\teff: effective temperature, \logg: surface gravity, and [M/H]: metallicity) and chemical abundances by comparing observed spectra to grids of synthetic spectra, minimizing $\chi^2$ in a multidimensional parameter space. In the 16th data release (DR16), a projected rotational velocity dimension was added to the the dwarf sub-grids (\logg~> 2.5 dex), and the \vsini~of dwarfs were determined by ASPCAP \citep{2020AJ....160..120J}. 

Gaia DR3 obtained the projected rotational velocities from the by-product of Extended Stellar Parametrizer for Hot Stars (ESP-HS), measuring the line broadening on the radial velocity spectrometer (RVS) spectra (8450 -- 8720 \AA). However, despite the relatively high resolution (R $\sim$ 11,500), ESP-HS values suffer from poor \vsini~-related information for hot stars such as OBA type due to the relatively limited wavelength range a\citep{2023A&A...674A..28F}.

Similar to other large sky surveys, the Large Sky Area Multi-Object Fiber Spectroscopic Telescope survey (LAMOST, \citealt{1996ApOpt..35.5155W, 2004ChJAA...4....1S, 2006astro.ph.12034Z, 2012RAA....12.1197C, 2012RAA....12..723Z}) developed the LAMOST stellar parameter pipeline (LASP) to automatically determine stellar parameters by analyzing Low-Resolution Survey (LRS) spectra, covering the wavelength range of 3800 -- 9000 \AA \ with a resolution of R $\sim$ 1800 \citep{2015RAA....15.1095L}. Since the LAMOST-II Medium-Resolution Spectroscopic (MRS) survey, the LASP was adapted to the MRS spectra (4950 -- 5350 \AA and 6300 -- 6800 \AA, R $\sim$ 7500). We updated LASP by adding a method to measure projected rotational velocity from the LAMOST spectra, using the synthetic spectra of PHOENIX \citep{2013A&A...553A...6H}. The updated LASP was developed to obtain a projected rotational velocity when other parameters have been determined. The updated LASP was applied to the LAMOST DR9 and here provides a catalog of 121,698 rotational velocities for MRS stars and for 80,108 LRS targets.

In this paper, we provided a thorough description of the \vsini~measurement in the updated-LASP. This paper is organized as follows. Section \ref{sec:data} details the data we used in this work. Section \ref{sec:method} described the method we adopted, its validation on APOGEE spectra and application to  LAMOST spectra. Sec. \ref{sec:results} display the results of this work and display external accuracy by comparing with catalogs. Finally, we summarize this work in Sec.~\ref{sec:summary}.
\section{Data} \label{sec:data}

\subsection {PHOENIX grids}
In this study, the version 16 of PHOENIX was adopted \citep{2013A&A...553A...6H}, which used a new equation of state and an up-to-date atomic and molecular line list. This allowed PHOENIX to produce synthetic spectra that match observations better than other synthetic libraries, especially for cool stars. 

The synthetic spectra cover the wavelength range from 500 \AA~to 5.5 $\mu$m in the optical and near IR, with resolution of R $ \sim$ 500,000. The grid coverage of \teff~is from 2300 K to 12,000 K, \logg~from 0.0 dex to 6.0 dex, and metallicity (\feh) from -4.0 dex to 1.0 dex. We did not take into account alpha-enhancement in this work, and selected the sub-grids of $[\alpha/Fe]$ = 0.0, which are  listed in Table \ref{tab1}.

\begin{table}
	\begin{center}
		\caption{Parameter Ranges of the PHOENIX Sub-grids Used in this work}\label{tab1}
		\begin{tabular}{lcc}
			\hline\noalign{\smallskip} \hline\noalign{\smallskip}
			& Range & Step size \\
			\hline\noalign{\smallskip}
			$T_{\rm eff}$ (K) & 2300 -- 7000 & 100 \\
                             & 7000 ---12,000 & 200 \\
			log \emph{g} (dex) & 0 -- 6.0 & 0.5 \\
			$[Fe/H]$ (dex) & -4.0 -- -2.0 & 1.0 \\
                                 & -2.0 -- +1.0 & 0.5 \\
				\noalign{\smallskip}\hline
		\end{tabular}
	\end{center}
\end{table}

\subsection{Preliminary Sample Selection} 
The LAMOST DR9 collected more than 3 million (1.4 million stars) MRS and 10 million (7.6 million stars) LRS spectra, of these, 0.8 million stars with 1.6 million MRS spectra were parameterized by the LASP and 5.1 million stars with 7 million LRS spectra were parameterized for LRS. LASP has the equivalent accuracies of stellar parameters on both MRS and LRS, which are about 100 K, 0.19 dex, and 0.10 dex for \teff, \logg, and \feh, respectively, in the effective temperature range of 4000 K < \teff~< 8500 K \citep{2015RAA....15.1095L}. This is because the MRS, despite its higher resolution, covers a very small wavelength range, from 4950 \AA \ to 5350 \AA \ for blue band, and 6300 \AA \ to 6800 \AA \ for red band. However, LASP can derive a higher precision radial velocity (RV) for MRS than LRS, with a precision of $\sim$ 1.5 \kms~compared to $\sim$ 5.0 \kms~\citep{2015RAA....15.1095L,2019ApJS..244...27W}.

Considering the wavelength coverage of the MRS, we selected stars with 5000 K < \teff~< 8500 K to determine \vsini~from the coadded spectra. We had no confidence in the parameters of stars with \teff~> 8500 K, because there's no balmer line in the blue part of the MRS spectrum that LASP used. For the LRS, the low resolving power (R $\sim$ 1800) allows us to detect only fast rotating stars, but fast rotating stars are rare in the late-type stars. Therefore, we abandoned the late-type stars and added a few early-type stars, which is in the effective temperature range of 7000 K < \teff~< 9000 K. 

\begin{figure}[ht!]
	\centering
	\includegraphics[width=0.8\linewidth]{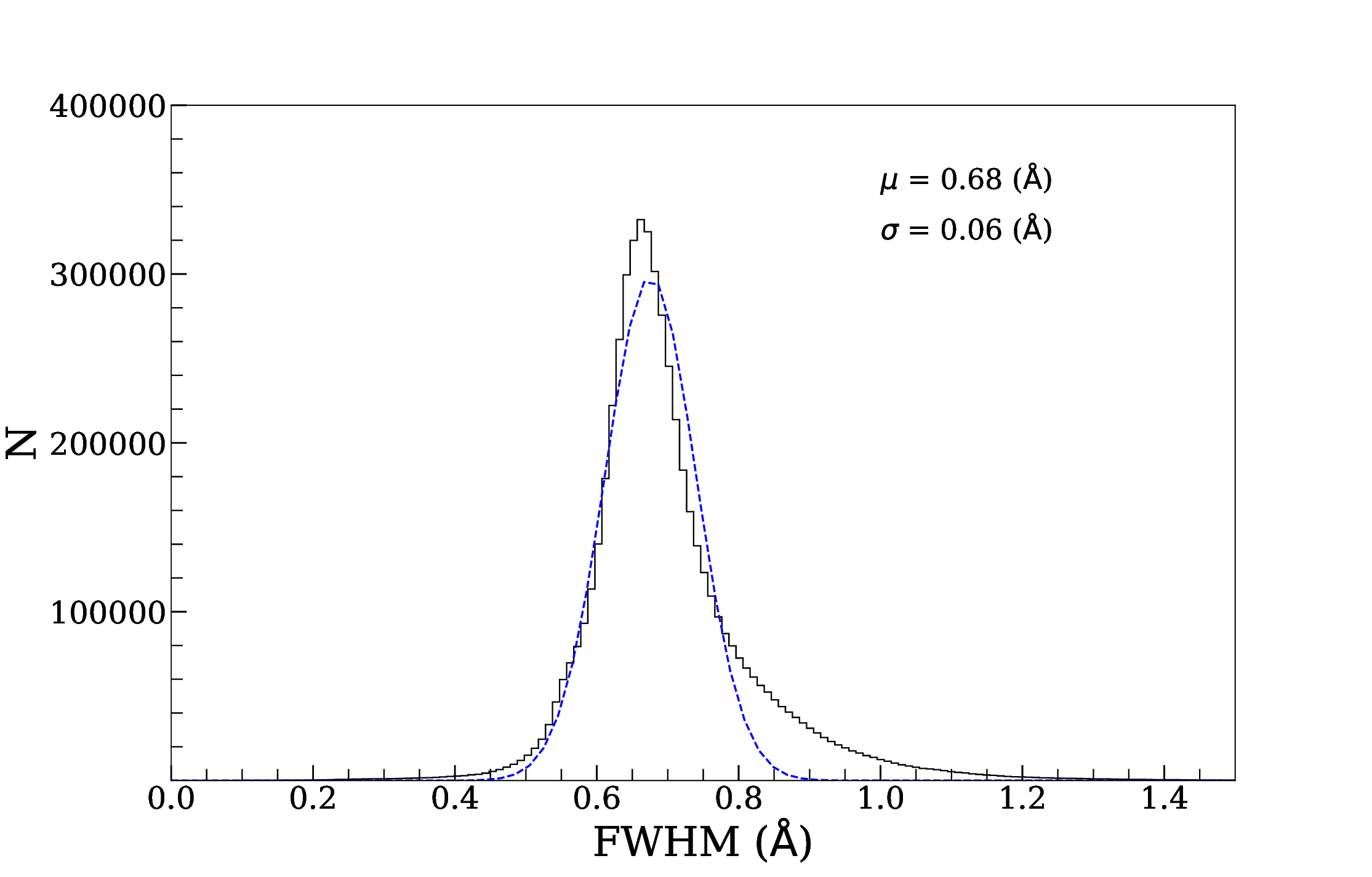}
	\caption{Histogram of the full width at half maximum (FWHM) of the thorium-argon wavelength calibration lines for MRS. The blue curve is the Gaussian fit to the FWHM distribution.
	        }
  \label{fig:arc_med}
\end{figure}

\begin{figure}[ht!]
	\centering
	\includegraphics[width=0.8\linewidth]{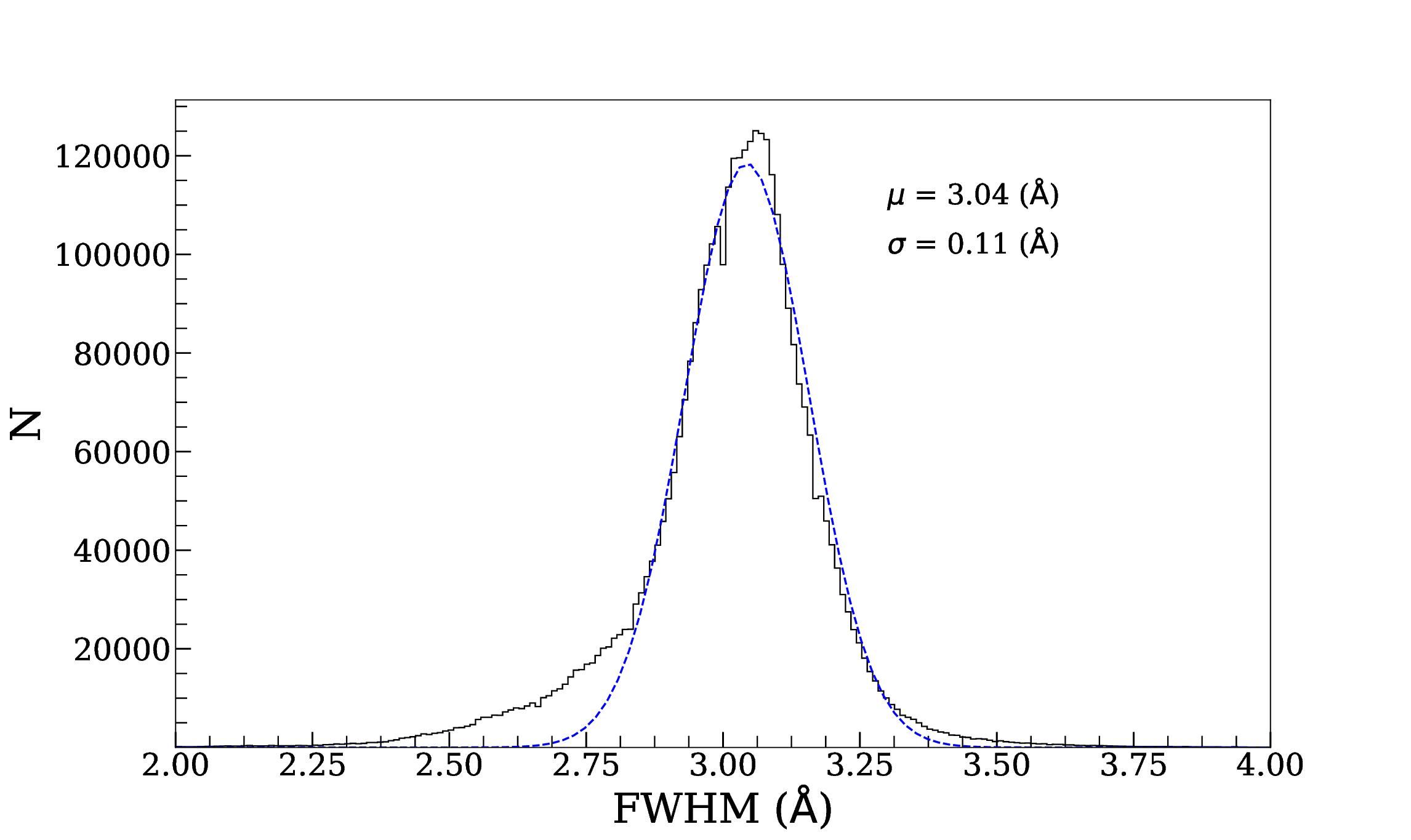}
	\caption{Histogram of the FWHM of the arc calibration lines for the LRS with best fit Gaussian plotted in blue. \ref{fig:arc_med}.	       }
  \label{fig:arc_low}
\end{figure}

In addition, both MRS and LRS observations have a nearly fixed instrumental profile with the mean full width at half maximum (FWHM) of $\mu$ = 0.68 \AA \ for MRS (see Figure \ref{fig:arc_med}) and the mean FWHM of $\mu$ = 3.04 \AA \ for LRS (see Figure \ref{fig:arc_low}), with a scatter of 0.06 \AA \ for MRS and 0.11 \AA \ for LRS. This indicates that resolutions of MRS and LRS vary with wavelength, such a resolution variation has an impact on the \vsini~, and we abandoned fibers with a FWHMs beyond $2\sigma$.

Non-single targets were also removed from samples previously selected by through cross-matching with Gaia DR3 \citep{2021A&A...649A...1G}, and catalog from \citealt{2023AJ....165..193W} and \citealt{2022MNRAS.513.5270P}, which offered nearby binaries, Stars with RUWE >= 1.4, and ipd\_gof\_harmonic\_amplitude < 0.1 or ipd\_frac\_multi\_peak < 10 \citep{2021MNRAS.506.2269E} were also cut out.

\section{Method} \label{sec:method}

In this work, we used $\chi^2$ minimization in the \vsini~space to derive the best-matched \vsini~under the constraints of other parameters.
 
\subsection{The determination method of \vsini} \label{sec:method1}

There are four steps for \vsini~determination, including generation of the reference spectrum by interpolating PHOENIX grids, matching of spectral resolution, convolution of broadening kernel, and $\chi^2$ minimization.

\begin{enumerate}

\item Generation of the synthetic reference spectrum. Since the atmospheric parameters have already been determined by LASP, we generated the reference spectrum by linearly interpolating the PHOENIX grids with atmospheric parameters around the LASP determined ones.

\item 
The resolution of generated reference spectrum was reduced to the resolution of observed spectrum.

\begin{figure}[ht!]
	\centering
	\includegraphics[width=0.8\linewidth]{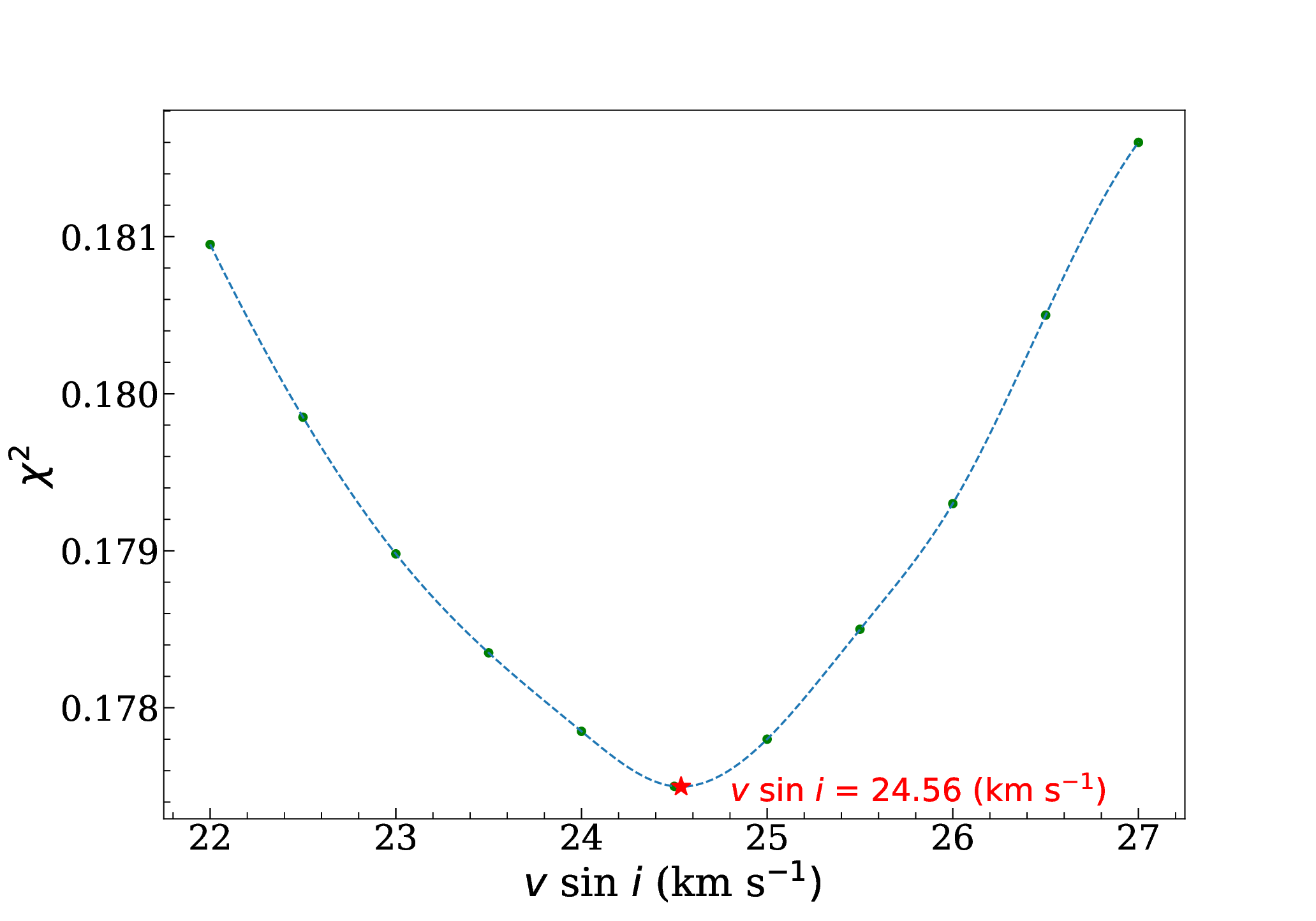}
	\caption{An example $\chi^2$ minimisation curve showing the fitting process. The blue dashed curve is the result of dense spline interpolation, and the red asterisk is the minimum $\chi^2$ point.
	        }
  \label{fig:chi}
\end{figure}

\item Convolution of broadening kernel. The broadening kernel was taken from \cite{Gray05} and given below:  

\begin{equation}
\delta=1-(\frac{v}{n})^2
\end{equation}
\label{eq:Leb1}
\begin{equation}
  G(\delta)=\frac{2(1-\varepsilon)\delta^{1/2}+\frac{1}{2}\pi\varepsilon\delta}{\pi n(1-\varepsilon/3)}
\label{eq:Leb2}
\end{equation}

Where $v$ is the rotational velocity \vsini~, $\varepsilon$ is the limb-darkening coefficient, n is the width of convolutional kernel, and its value is resolution-dependent. We adopted $\varepsilon$ = 0.6, n = 7 for MRS and n = 75 for LRS, we produced 700 spectra with different \vsini~through convolving the reference spectrum with different rotational kernels from \vsini~= 1 \kms~to 350 \kms~(step = 0.5 \kms).

\item $\chi^2$ minimization. We minimized $\chi^2$ by comparing the observed spectrum to the convolved reference spectra. Then we fitted the 11 points of (\vsini~, $\chi^2$) which have the minimum $\chi^2$ in the middle (see the blue points in Figure \ref{fig:chi}), using a dense spline interpolation method instead of a Gaussian to avoid fitting failure \citep{2019ApJS..240...10D}. The minimum $\chi^2$ point of the dense interpolation corresponds to the projected rotational velocity of the observed spectrum (see the red star in Figure \ref{fig:chi}).

\end{enumerate}

\subsection{Validation on the APOGEE spectra}

\begin{figure*}[ht!]
	\centering
	\includegraphics[width=0.5\linewidth]{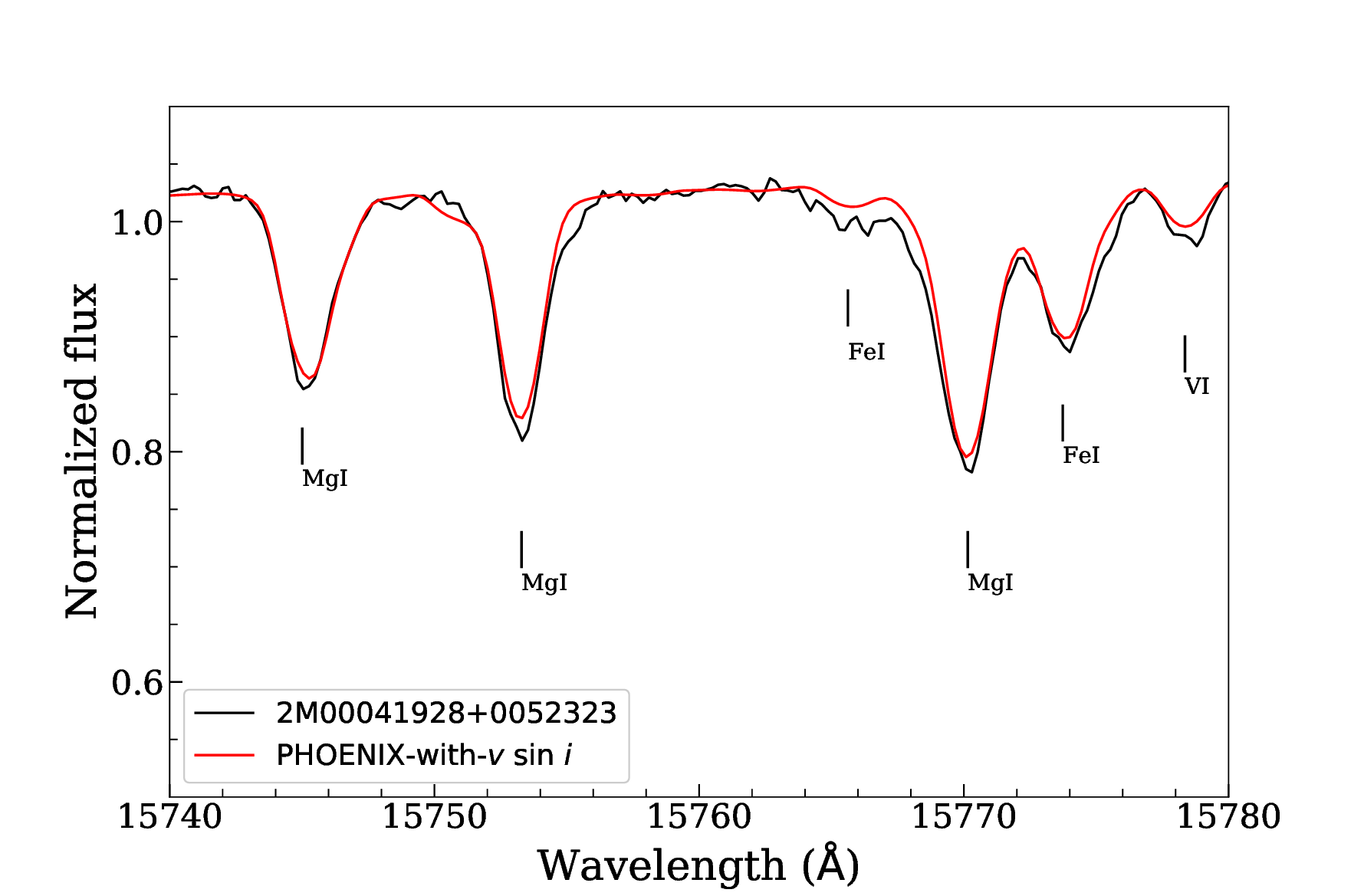}
	\caption{An example APOGEE spectrum is plotted in black along with a best fit model spectrum in red. The ASPCAP parameters of this target are \teff~= 6188.34 K, \logg~= 4.2 dex, and [Fe/H] = -0.01 dex. The reference spectrum from PHOENIX includes  rotational velocity broadening of 13.3 \kms.}
  \label{fig:apogee_spec}
\end{figure*}

To validate our method, we performed the method on APOGEE spectra with a resolution of R $\sim$ 22,500. The APOGEE spectra included three bands, [15,100 -- 15,799 \AA], [15,867 -- 16,424 \AA], and [16,484 -- 17,000 \AA], and we used the spectrum of 15,740 -- 15,780 \AA, where four strong isolated lines exist (see Figure \ref{fig:apogee_spec}), to measure \vsini~.

Because of the resolution of APOGEE spectra, stars rotating faster than 8 \kms~can be detected \citep{2016A&A...594A..39F}. We selected the sample from APOGEE DR17 catalog with the ASPCAP \teff~> 5000 K and a \vsini~greater than 8 \kms, and finally a sample of 27,000 spectra were selected..

\begin{figure*}[ht!]
	\centering
	\includegraphics[width=\linewidth]{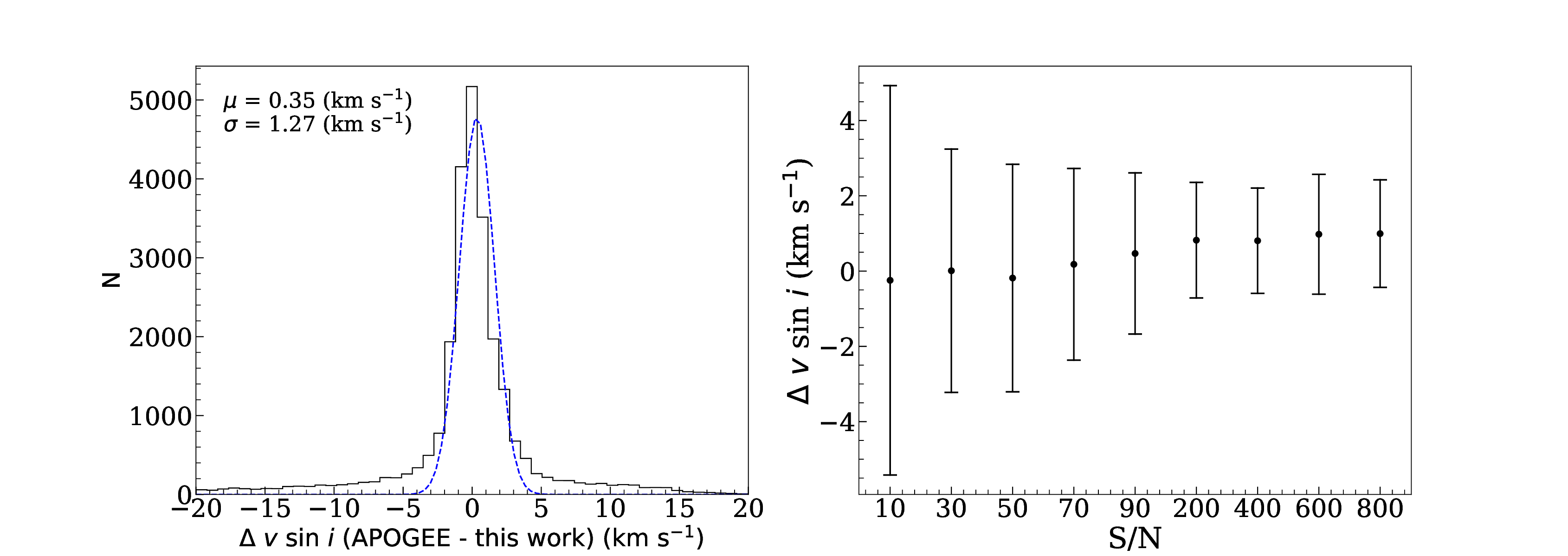}
	\caption{Comparison of our \vsini~results with those from  ASPCAP. The left panel shows the histograms of differences between results of this method and those of ASPCAP along with Gaussian fit in blue. The right panel shows the distribution in  \vsini~values as a function of signal-to-noise (S/N).}
  \label{fig:apogee_com}
\end{figure*}

\begin{figure}[h!]
	\centering
	\includegraphics[width=0.9\linewidth]{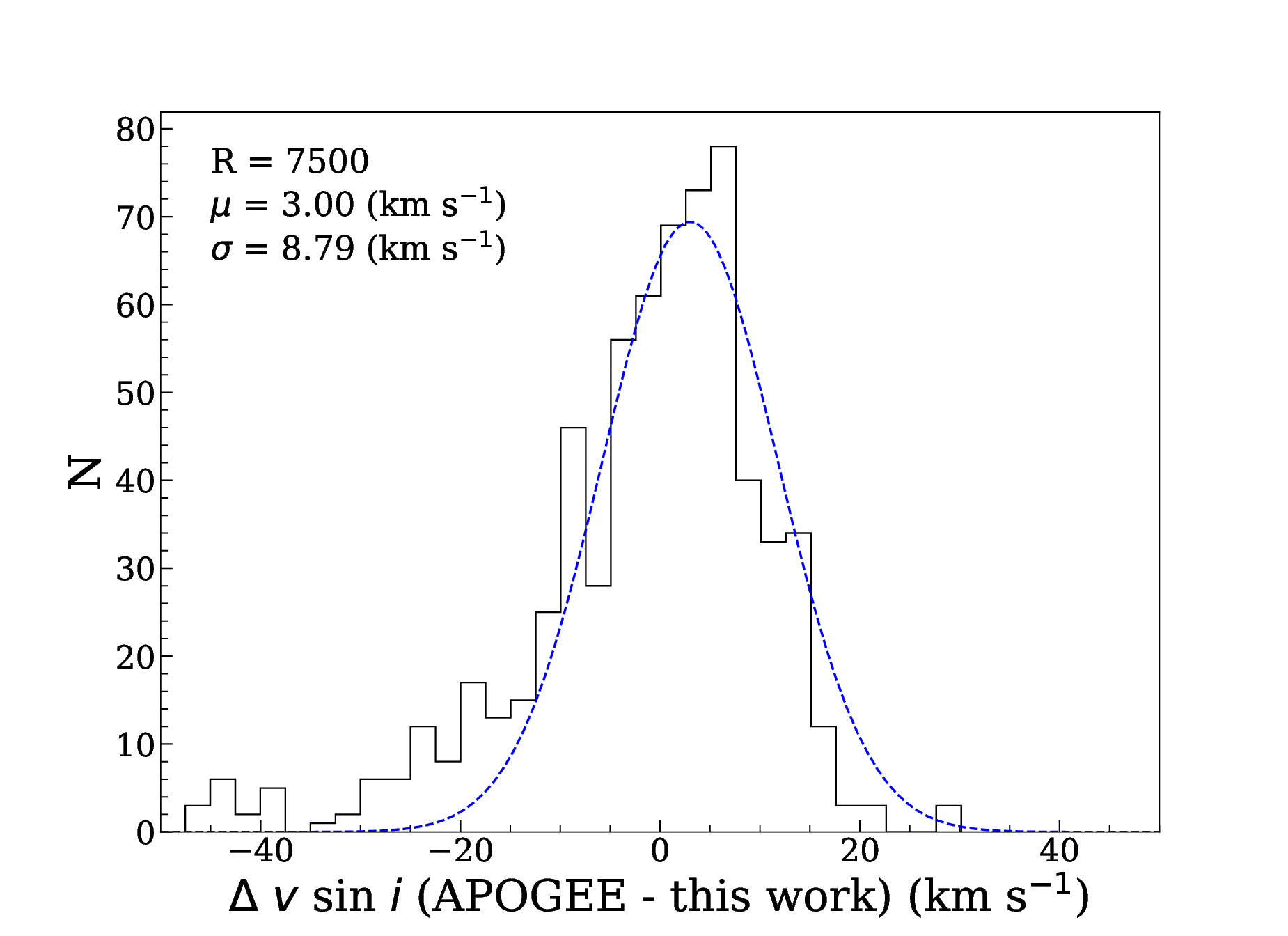}
	\caption{The histogram of the difference between our results and those of ASPCAP is shown along with a Gaussian fit in blue.
	        }
  \label{fig:apogee_com_r1}
\end{figure}

\begin{figure*}[ht!]
	\centering
	\includegraphics[width=0.8\linewidth]{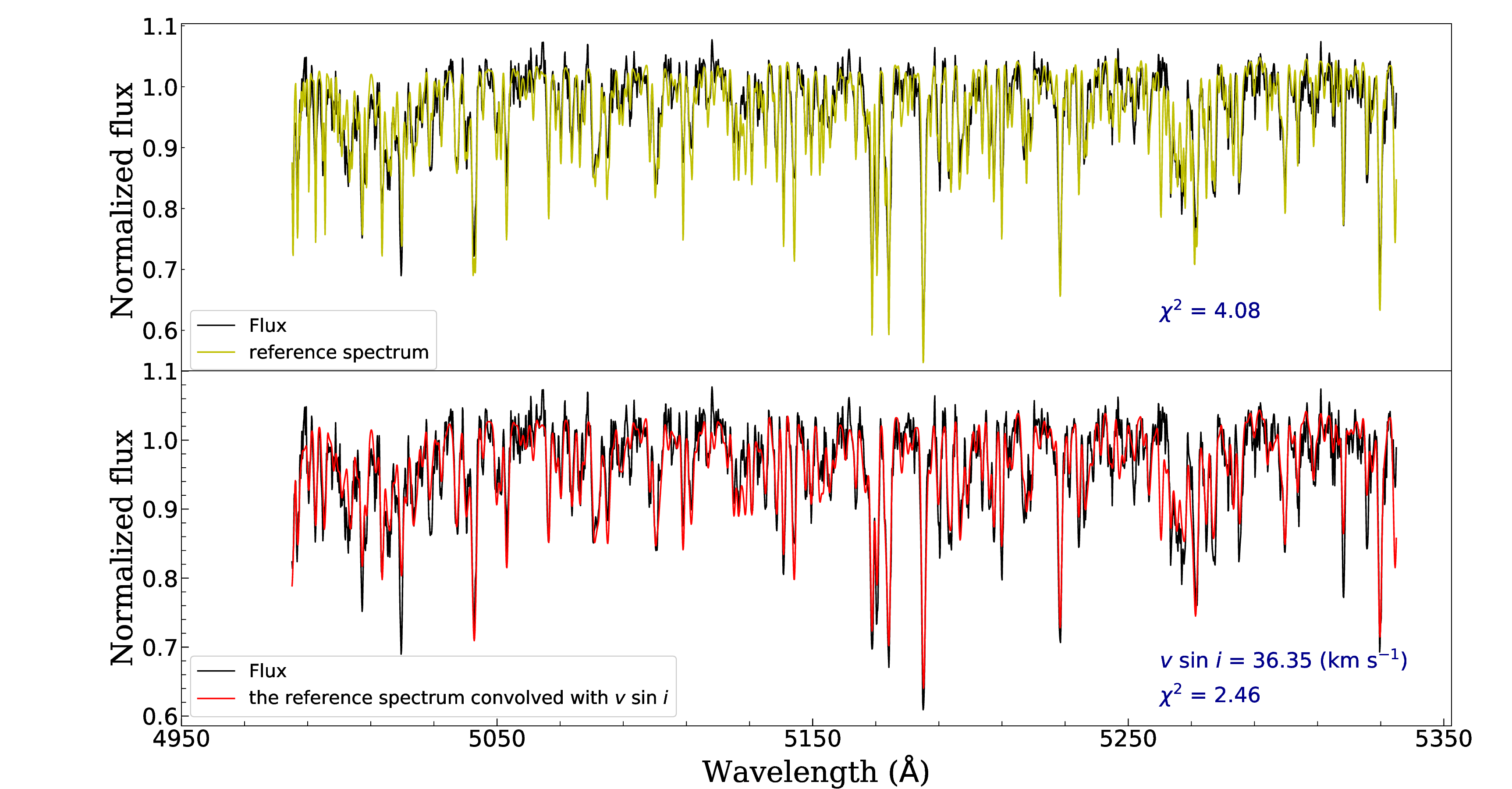}
	\caption{A comparison between an MRS spectrum in black with its best fit model spectrum in yellow and red. The upper plot shows the model spectrum in yellow prior to broadening. The lower plot shows the synthetic spectrum following convolution with a suitable \vsini~. 
	        }
  \label{fig:mrs-spec}
\end{figure*}

We determined \vsini~for the 27,000 spectra and compared our results with the \vsini~of ASPCAP, which is shown in Figure \ref{fig:apogee_com}. The results of our method are consistent with those of ASPCAP, with a scatter of 1.27 \kms. To inspect the influence of S/N on the \vsini~measurement, we calculated the mean and the standard deviation of the difference between our results and those of ASPCAP under different S/Ns (see the right panel in Figure \ref{fig:apogee_com}), and noticed that the \vsini~measurement is greatly affected by the S/N for S/N < 30, while less affected by S/N > 50.

We degraded the resolution of APOGEE spectra to R = 7500, to evaluate the performance of the spectra with the same resolution of LAMOST, and the result shows that the LAMOST MRS resolution allows us to detect only stars rotating faster than 27 \kms. We further excluded samples with the ASPCAP \vsini~less than 27 \kms~in the evaluation, Figure \ref{fig:apogee_com_r1} shows the comparison of our results to the ASPCAP \vsini~, and the difference has a scatter of 8.79 \kms. 

\subsection{Application to LAMOST} \label{sec:application}

In this work, we estimated \vsini~for both MRS and LRS in LAMOST DR9, using the blue band spectra with 4950 -- 5300 \AA \ for MRS and 4200 -- 5700 \AA \ for LRS. 
The application of \vsini~measurements to both surveys will be described below.

\subsubsection{Application to the MRS}
Consistent with the spectral band used in LASP, we also used the blue arm spectra (4950 -- 5300 \AA) to measure \vsini, without using the red arm because of the possible existence of strong H$\alpha$ emission \citep{2019ApJS..244...27W}. We adopted the stellar parameters of LASP, including \teff, \logg, \feh, and RV, and interpolated the synthetic spectra of the PHOENIX grids based on the LASP \teff, \logg, and \feh. Then, we degraded the resolution of reference spectrum to the resolution of MRS, and subtracted pseudocontinua of both the synthetic and the observed spectrum by polynomial fitting. We added the broadening kernel that was given in Section \ref{sec:method1} to the synthetic spectrum (see Figure \ref{fig:mrs-spec}), and \vsini~is determined by minimizing $\chi^2$.

\subsubsection{Application to the LRS}
For the LRS, we used the spectra of 4200 -- 5700 \AA \ to determine \vsini. We didn't use the H$\alpha$ (6520 -- 6595 \AA) band, because of its instability and that an RV offset of $\sim$ 7 \kms~exists in this band \citep{2019ApJS..240...10D}. We didn't take into account the Ca II triplet (8400 -- 8700 \AA), because the lines are weak for hot stars. Otherwise the measurement procedure for the LRS was consistent with that for the MRS.

\section{The Results} \label{sec:results}



\subsection{Precision} \label{sec:precision}
We estimated \vsini~precision from the \vsini~measurements of the multiple observations for the same stars \citep{2021RAA....21..202D}. The targets with more than three measurements were selected for the calculation, and the statistical estimator used to assess the precision is:

\begin{equation}
   \epsilon=\sqrt{(n/(n-1))} \times (vsini_{\rm i} - \overline{vsini})
\label{eq:Leb3}
\end{equation}

where $\emph{n}$ is the observation times of the same star, $vsini_{\rm i}$ is the \vsini~of the ${\rm i_{\rm th}}$ observation, \emph{i} = 1, 2, ..., $\emph{n}$, and $\overline{vsini} = ~\frac{1}{n} \sum_{i=1}^{n}vsini_{\rm i}$.

\begin{figure*}[ht!]
	\centering
        \includegraphics[width=0.45\linewidth]{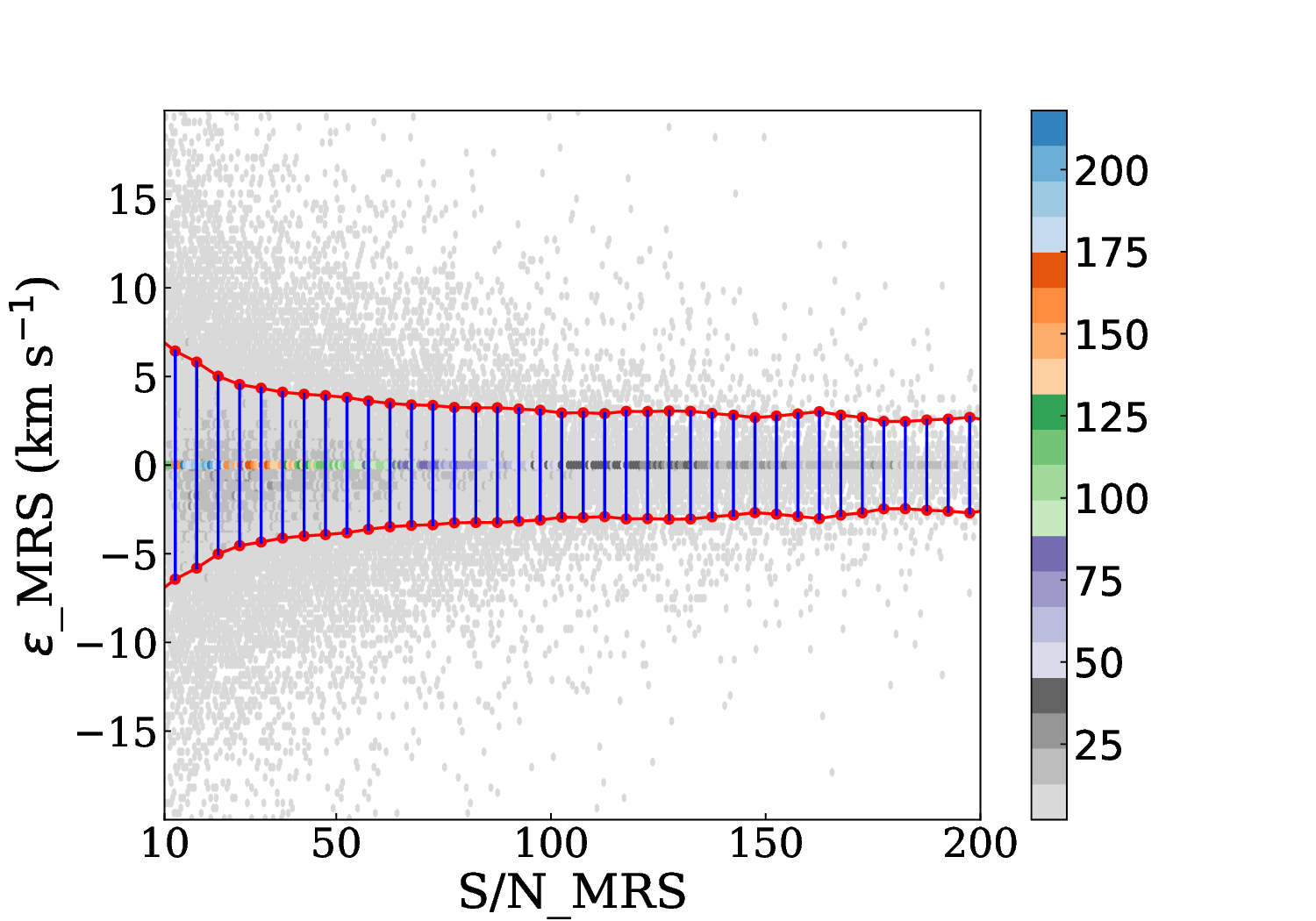}
        \includegraphics[width=0.45\linewidth]{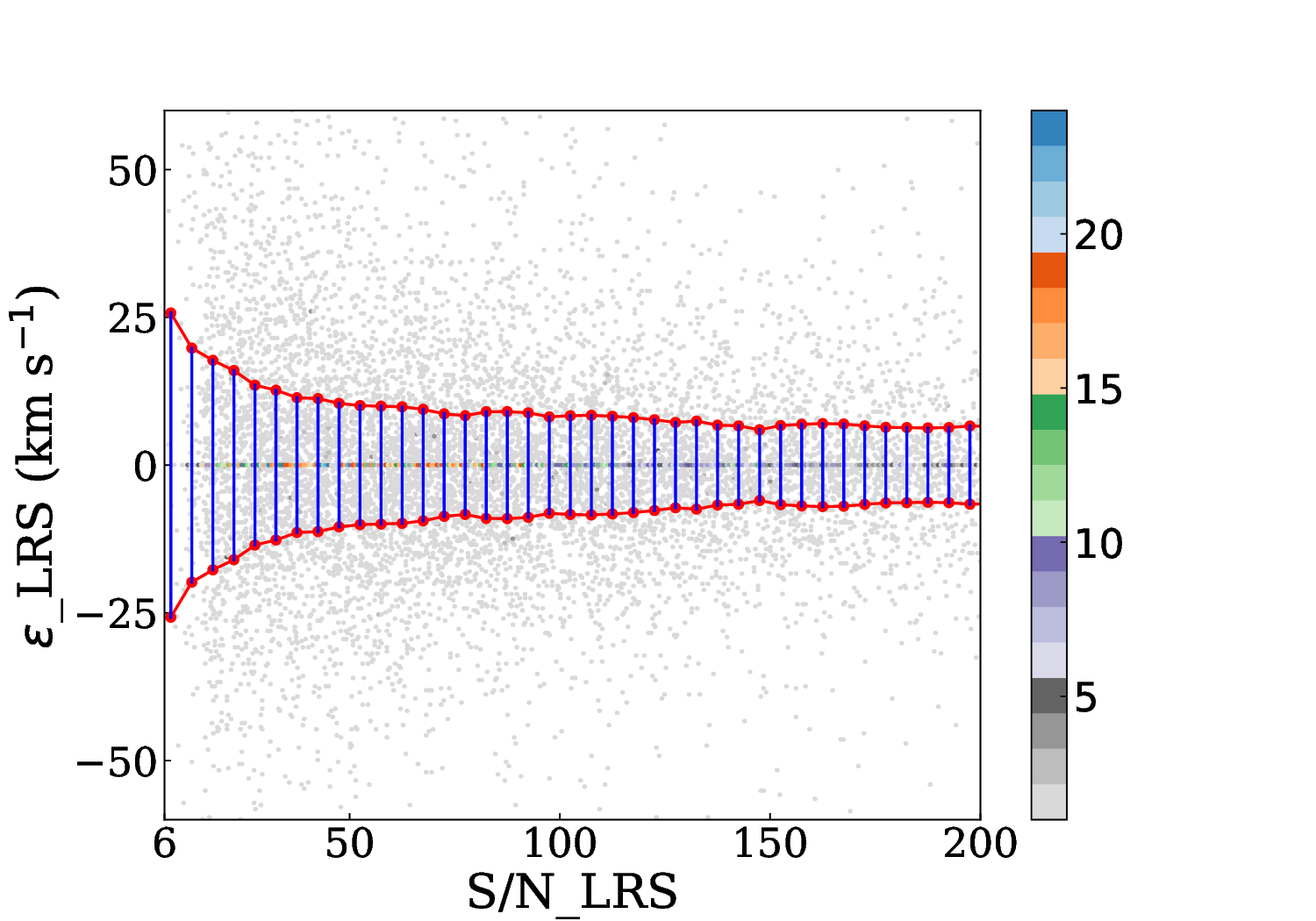}
        
	\caption{The distribution of \vsini~precision at different S/Ns for MRS (left) and LRS (right). The color contours indicate the number of stars, and the red curves are the spline fits to the 1$\sigma$ uncertainties of $\epsilon$ with a S/N step of 5.}
	\label{fig:esp}
\end{figure*}


Figure \ref{fig:esp} shows the distribution of \vsini~precision at different S/Ns for MRS and LRS. For both of them, \vsini~precisions are highly dependent on the S/Ns when S/N < 50, while a fixed level of about 4.0 \kms~as S/N > 50 is obtained for MRS and about 10.0 \kms~when S/N > 50 for LRS.

\begin{figure*}[hb!]
	\centering
	\includegraphics[width=1.0\linewidth]{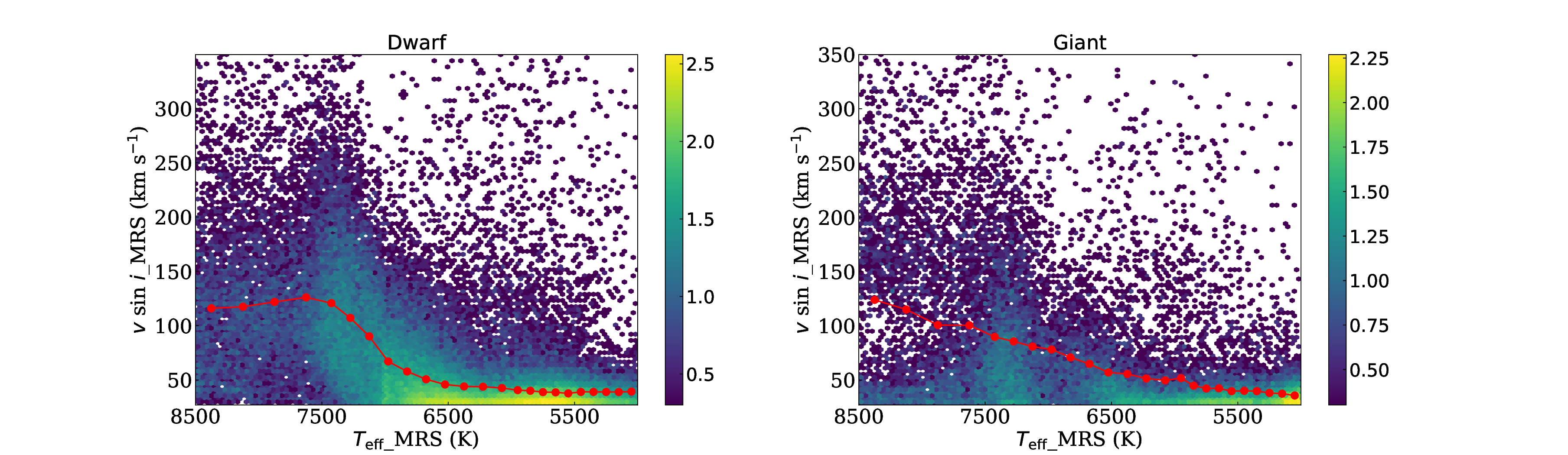} 
 	
	\caption{\textbf{The \vsini~ -- \teff~ diagrams for dwarf stars (left) and giant stars (right), color-coded by the log10 (number). The red curves are the spline fits to the average \vsini~ within non-uniform \teff~ bins. }}
	\label{fig:mrs1}
\end{figure*}

\begin{figure*}[hb!]
	\centering
	\includegraphics[width=0.7\linewidth]{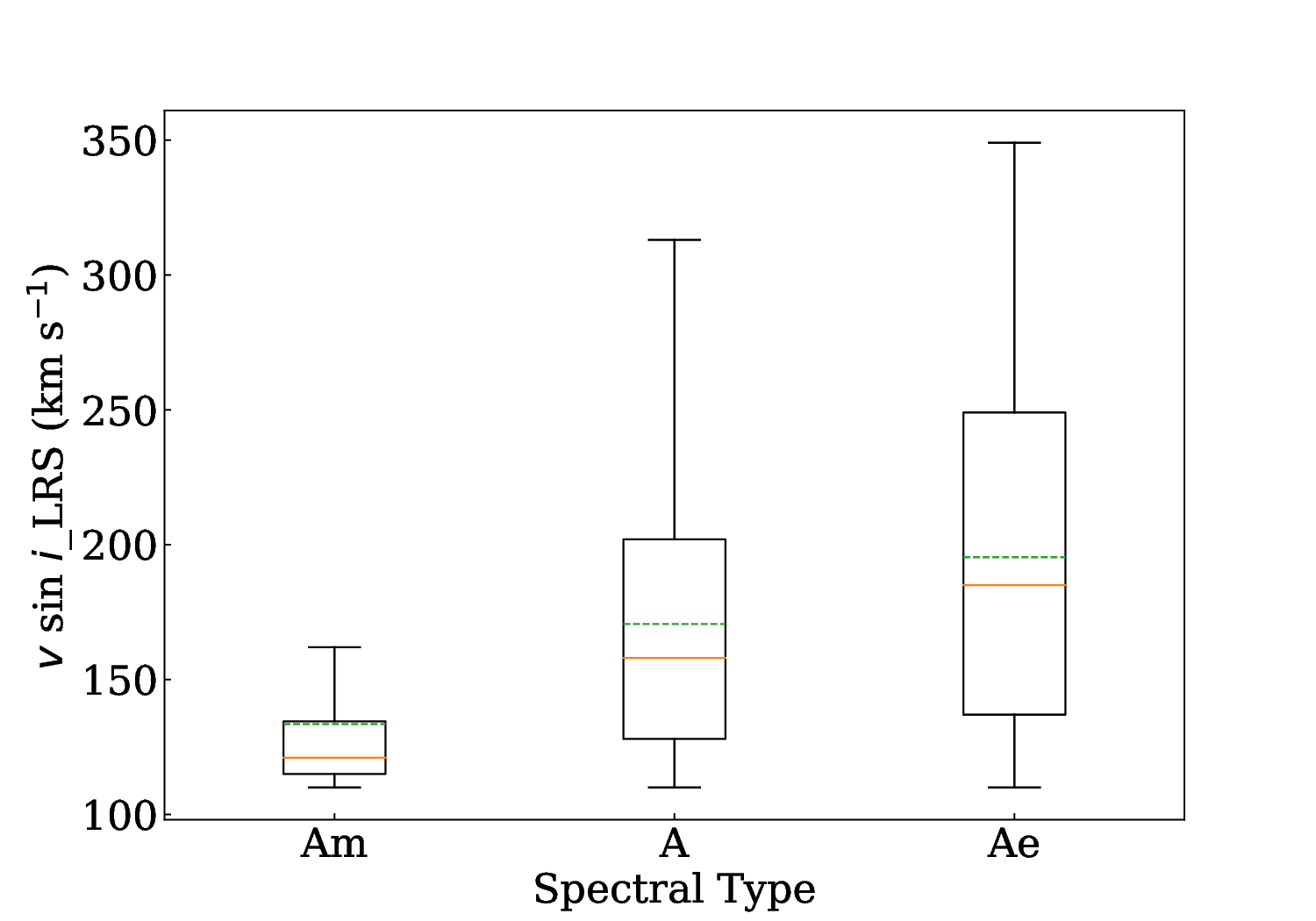} 
 	
	\caption{\textbf{The box extends from the lower to upper quartile values of the LRS \vsini~ of Am, A and Ae stars, with green dashed (orange) line at the mean (median). The vertical line extend from the box to show the \vsini~range. }}
	\label{fig:ae}
\end{figure*}

\begin{deluxetable*}{ccc ccc ccc}[ht]
	\tablenum{2}
	\tablewidth{700pt}
	\tabletypesize{\scriptsize}
	\tablecaption{Parameters of LAMOST MRS \vsini~CATALOG. 
		\label{tab:parameters} }
	\tablehead{
		\colhead{UID$^a$} & \colhead{R.A.} & \colhead{Dec.} & \colhead{S/N} & \colhead{\teff} & \colhead{\logg} & \colhead{\feh} & \colhead{\vsini$^b$} & \colhead{\vsini\_err$^c$} \\
		\colhead{ } & \multicolumn{2}{c}{(deg)} & \colhead{ } & \colhead{(K)} & \multicolumn{2}{c}{(dex)} & \multicolumn{2}{c}{(\kms)} }
	
    \startdata 
	G16865597923189&22.4280930&54.7624245&40&7662&3.98&0.14&224&2\\
        G16865927110078&26.0557060&54.3317642&42&7850&3.91&0.02&90&4\\
        G17457893282113&83.7000580&42.1455650&266&8135&4.02&-0.19&130&1\\
        G17462346236306&75.8974991&48.2667503&114&7570&4.03&-0.05&113&1\\
	G13295087724001&340.1962585&24.3700695&150&7473&3.91&0.27&45&4\\
	G13292221927576&344.4146423&32.3737564&181&7919&4.09&0.01&92&2\\
	G13296262504355&340.7543030&27.2950096&178&7527&4.15&-0.16&262&1\\
        G13302953250167&356.0117493&29.4897861&158&7911&4.14&-0.14&118&1\\
        G13307681691870&351.3109436&24.0695324&306&7672&4.01&0.21&94&2\\
        G13278647737207&354.9405212&39.4682617&370&7702&3.94&0.07&58&2\\ 
        \\ \hline
    \enddata    
\tablecomments{LAMOST LRS \vsini~CATALOG has the same format, so we didn't display it. \\
               $^a$: LAMOST unique target ID\\
               $^b$: projected rotational velocity derived with the method in this work\\
               $^c$: error estimated by the method laid out in sub-section \ref{sec:precision}
               }
\end{deluxetable*}

\subsection{Catalog}

We construct two \vsini~catalogs for LAMOST MRS and LRS, i.e., LAMOST MRS \vsini~CATALOG and LAMOST LRS \vsini~CATALOG, which include 121,698 stars (221,770 spectra) for MRS and  80,108 stars (102,598 spectra) for LRS respectively.The format for both MRS and LRS are same and here we display some example lines of the  MRS \vsini~CATALOG in Table \ref{tab:parameters}. The two full catalogs can be accessed via the link https://nadc.china-vo.org/res/r101317.  For LAMOST MRS \vsini~CATALOG, we didn't have confidence in \vsini~< 27 \kms~as mentioned in \cite{2016A&A...594A..39F}, stars with \vsini~< 27 \kms~were removed, and for LAMOST LRS \vsini~CATALOG, stars with \vsini~< 110 \kms~ were removed for the same reason. Finally, 121,698 stars are included in LAMOST MRS \vsini~CATALOG, 80,108 stars for LAMOST LRS \vsini~CATALOG, and for stars with multiple observations, \vsini~and uncertainties calculated from the spectrum with the highest S/N were given in the two catalogs.  



The behaviors of rotation along the main sequence and giants were studied by \citet{1967ApJ...150..551K} \& \citet{1989ApJ...347.1021G}, stars with near solar abundances (-0.5 $\leq$\ \feh~$\leq$ 0.5 dex) in the LAMOST MRS \vsini~CATALOG were separated into dwarfs as \logg~$\ge$ 4.0 dex and giants as \logg~< 4.0 dex to inspect the behaviors in it. We didn't take into account the LAMOST LRS \vsini~CATALOG because of its limited temperature range of 7000 K < \teff~< 9000 K. Figure \ref{fig:mrs1} shows the distribution of \vsini~at different temperatures, and it sharply decreases at \teff~$\sim$ 7000 K for dwarfs, which is consistent with the ‘Kraft break’ around F0, and the average rotational velocity of stars with \teff~> 7500 K (see the red curve in left panel of Figure \ref{fig:mrs1}) is greater than 120 \kms, which is similar as the result in \cite{1982PASP...94..271F} . For giants, the break occurs at \teff~$\sim$ 6500 K, which is consistent with the rotation behavior in \citet{1979A&A....74...38B} \& \cite{1989ApJ...347.1021G}, and their average \vsini~is larger than 50 \kms, which represents giant stars have a gentle \vsini~transition (as shown by the red curve in the right panel of Figure \ref{fig:mrs1}). 

As mentioned in introduction, compared to typical A type star, Am and Ae are slower and faster rotators, respectively.  We cross-matched the \vsini~LRS CATALOG with the catalogs of \cite{2022ApJS..259...63S} (Am stars), ‘LAMOST Spectral Index of A type Stars’, and \cite{2022ApJS..259...38Z} (Ae stars), and obtained 387 Am, 49,107 typical A-type (removed Am and Ae) and 2215 Ae stars. The rotational velocities of the three type stars are shown in Figure \ref{fig:ae}, and their average \vsini~are 121 \kms, 158 \kms and 185 \kms, respectively, which are agreed with the results in \cite{1982PASP...94..271F}. We didn't consider the MRS \vsini~CATALOG here, because \teff~in this catalog are all lower than 8500 K, which means no hot A-type stars can be used.
  
\subsection{Comparison of MRS and LRS}
\begin{figure*}[ht!]
	\centering
	\includegraphics[width=0.8\linewidth]{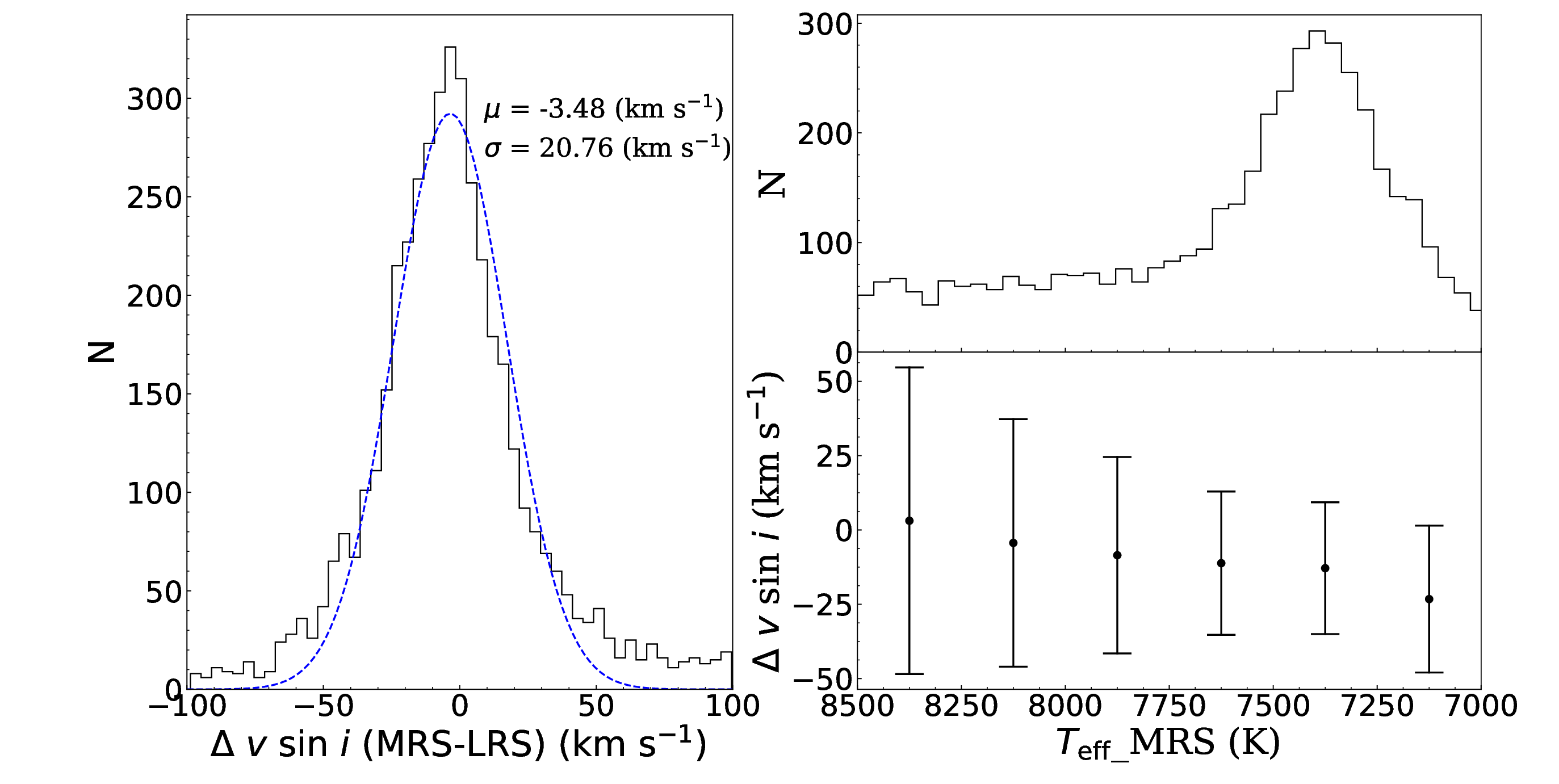}
	\caption{Comparison of the results between MRS and LRS data. The distribution of \vsini~differences between the MRS and LRS is shown in the left panel. The upper right panel shows the number of \vsini~values from the MRS LASP as a function of  \teff. The distribution of \vsini~differences between the MRS and LRS at different temperatures is shown in the lower right panel.
	        }
  \label{fig:mrs-lrs}
\end{figure*}

We cross-matched LAMOST MRS \vsini~CATALOG with LAMOST LRS \vsini~CATALOG finding 6000 stars in common, which are all fast rotating stars because the LAMOST LRS \vsini~CATALOG only provides \vsini~> 110 \kms. The difference distribution of \vsini~for these stars is shown in Figure \ref{fig:mrs-lrs}, with a mean of 3.45 \kms~and a standard deviation of 20.76 \kms~, and the relative error of difference is less than 2\%.

The \vsini~differences at different temperatures is shown in Figure \ref{fig:mrs-lrs} (right panel), and we found that the differences become larger at \teff~> 8000 K when compared to 7000 < \teff~< 8000 K, which is because the higher the temperature, the weaker the spectral lines of MRS blue arm spectra (4950 -- 5300 \AA). We noted a slight downward trend in the bottom right panel of Figure \ref{fig:mrs-lrs}, this is probably because the mixing of metal lines intensifies with the decrease of temperature for LRS, resulting in a whole overestimate of rotational velocity, that's why we didn't measure \vsini~for stars with \teff~< 7000 K.

\subsection{Comparison with other catalogs}

To verify the reliability of the \vsini~obtained from our method, we compared our results to other catalogs, including APOGEE DR17, Gaia DR3, and SUN (\citealt{2021ApJS..257...22S}, hereafter SUN), where APOGEE DR17 focus on F, G, K type stars, while Gaia DR3 and SUN focus on early type stars (F- and A-type). In order to reduce the uncertainty introduced by the stellar parameters, we selected the stars with temperature difference less than 500 K, and the differences of both \logg~and \feh~less than 1.0 dex.

\subsubsection{Comparison with APOGEE DR17}

\begin{figure}[ht!]
	\centering
        \includegraphics[width=0.8\linewidth]{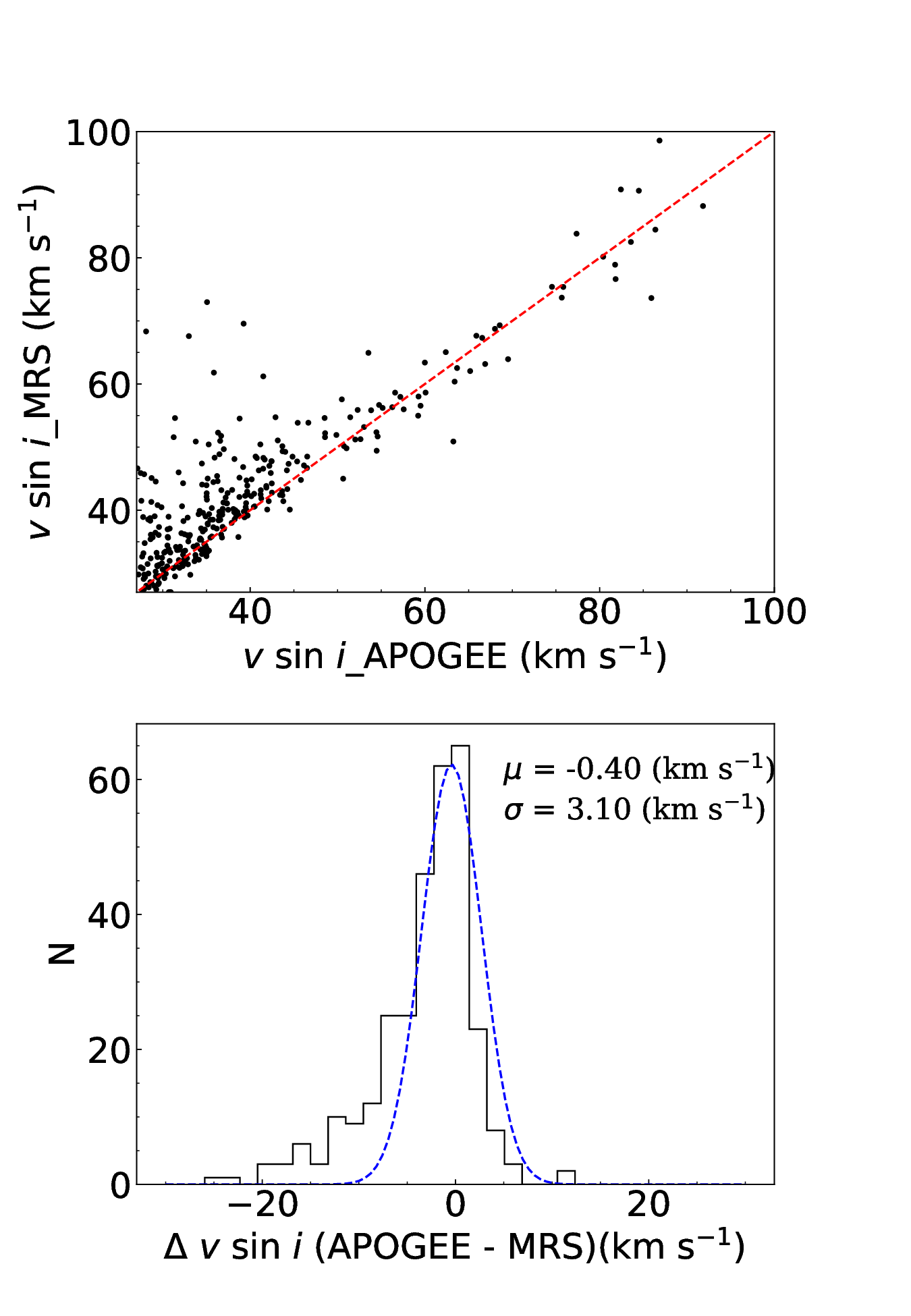}
	\caption{Comparison of MRS \vsini~results in this work to APOGEE DR17 with the red line showing the one-to-one line. The lower plot shows the histogram of \vsini~differences between the two catalogs with gaussian fit in blue. 
	        }
  \label{fig:vs-apogee1}
\end{figure}

Projected rotational velocities released in APOGEE DR17 are less than 100 \kms, which have no overlap with the LAMOST LRS \vsini~CATALOG, thus we only compare LAMOST MRS \vsini~CATALOG with APOGEE DR17, and obtain 311 common stars. Figure \ref{fig:vs-apogee1} shows the comparison results, and we notice that \vsini~given in this work are consistent with ASPCAP results, with a small offset of 0.40 \kms, and a scatter of 3.10 \kms.

\begin{figure}[ht!]
	\centering
        \includegraphics[width=0.8\linewidth]{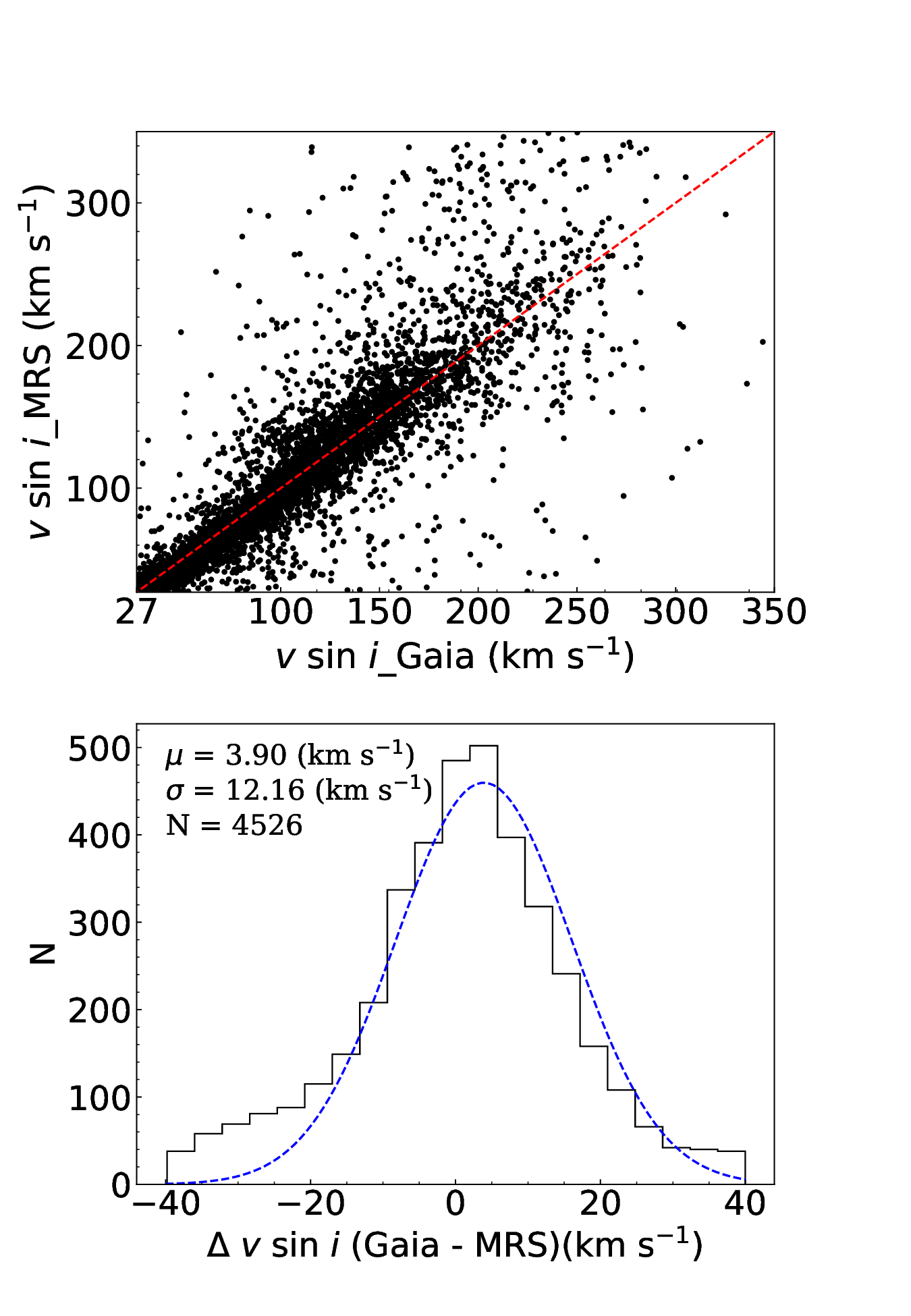}        
	\caption{Comparison of MRS \vsini~results in this work to Gaia DR3 following the conventions of Figure \ref{fig:vs-apogee1}.
        }
  \label{fig:mrs-gaia}
\end{figure}

\begin{figure}[hb!]
	\centering
        \includegraphics[width=0.8\linewidth]{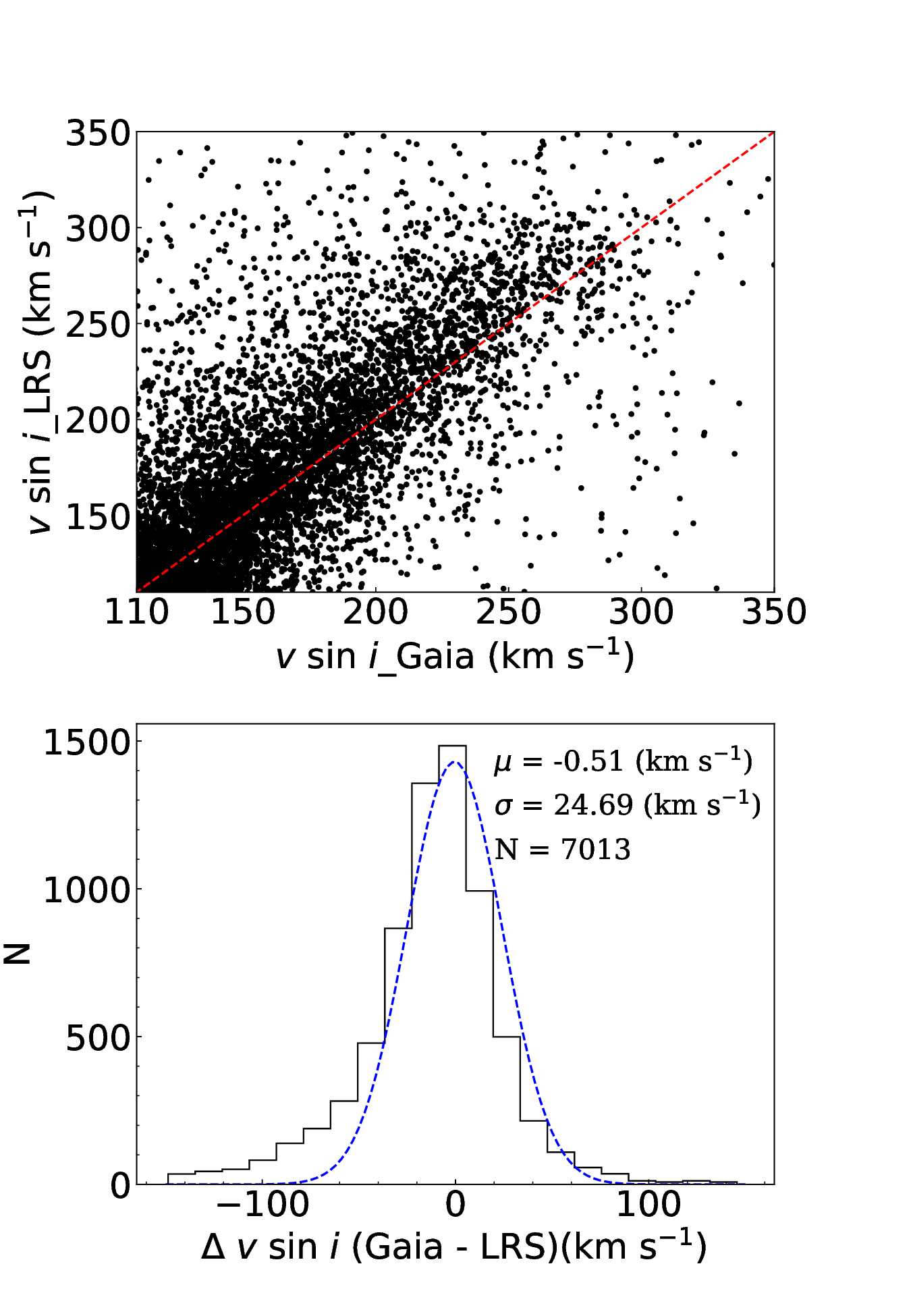}        
	\caption{ Comparison of the LRS \vsini~to Gaia DR3 following the conventions of Figure \ref{fig:vs-apogee1}.
        }
  \label{fig:lrs-gaia}
\end{figure}

\begin{figure}[ht!]
	\centering
        \includegraphics[width=0.8\linewidth]{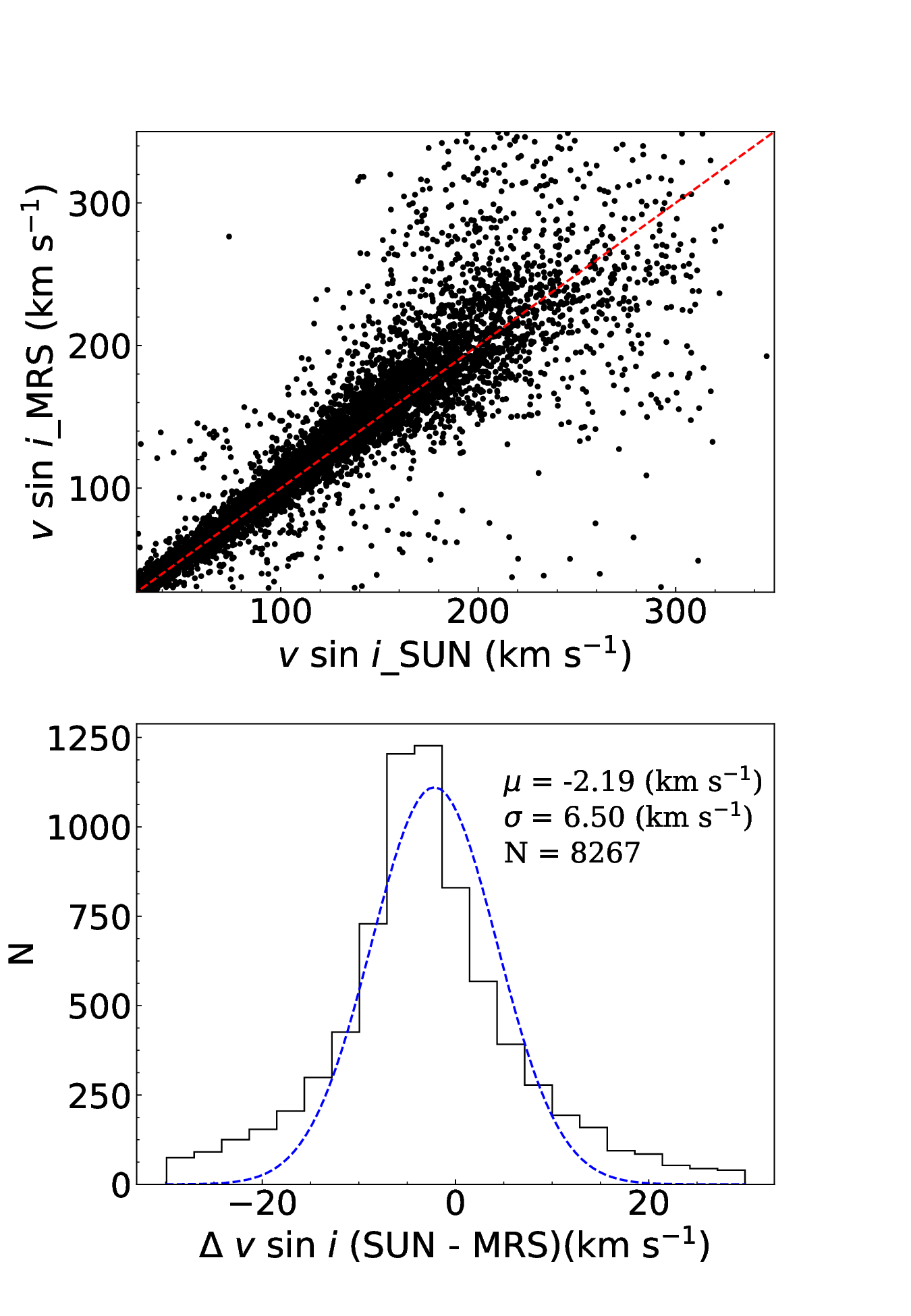}        
	\caption{ Comparison of the MRS \vsini~to SUN following the conventions of Figure \ref{fig:vs-apogee1}.
        }
  \label{fig:mrs-sun1}
\end{figure}

\begin{figure}[hb!]
	\centering	
        \includegraphics[width=0.8\linewidth]{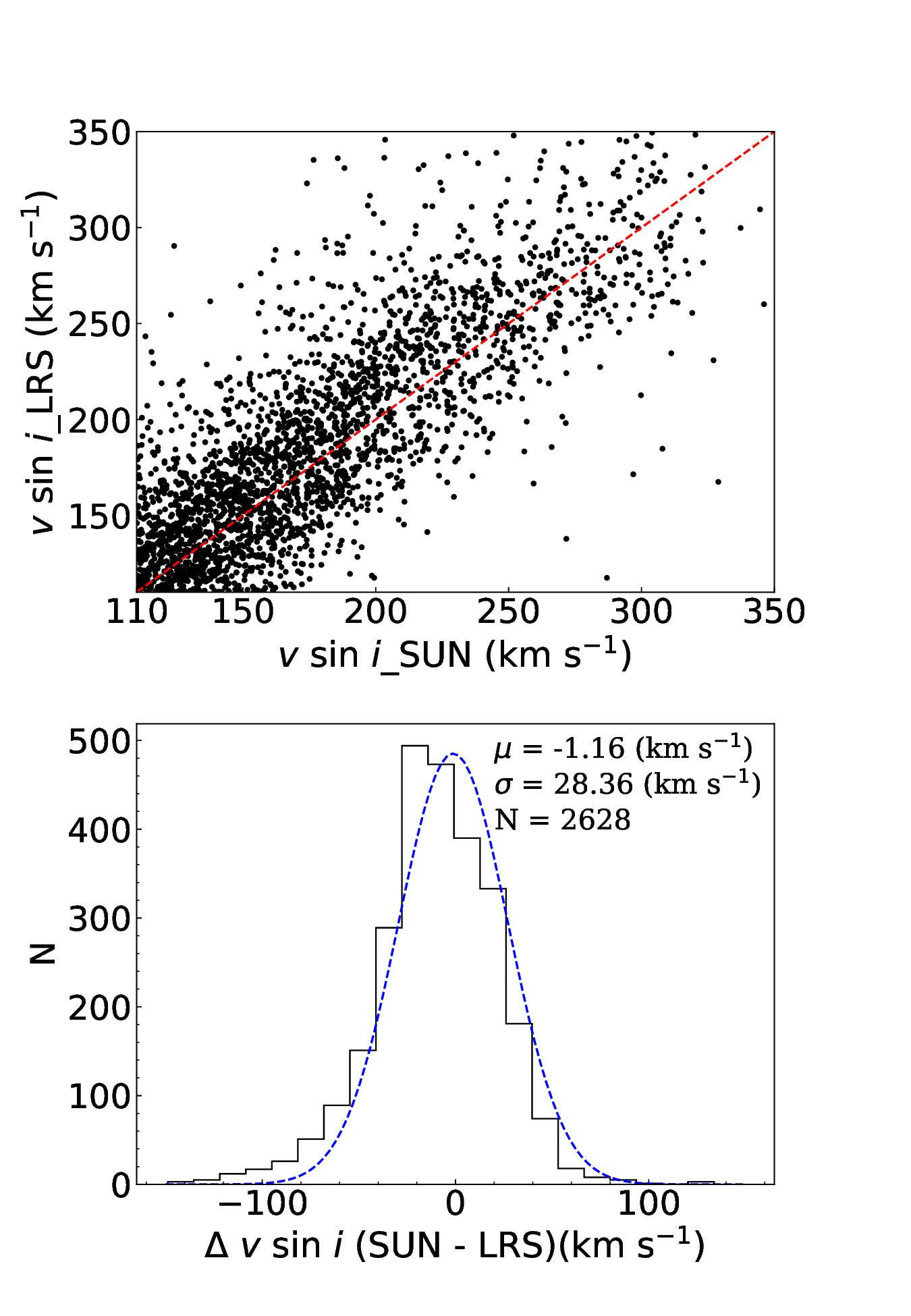}
	\caption{ Comparison of the LRS \vsini~to SUN following the conventions of Figure \ref{fig:vs-apogee1}.
        }
  \label{fig:lrs-sun}
\end{figure}

\subsubsection{Comparison with Gaia DR3}

Gaia is a space mission of the European Space Agency (ESA) started in 2013 (\citealt{2016A&A...595A...1G}), and the first data was released in 2016 \citealt{2016A&A...585A..93R,2016A&A...595A.133C}) with the aim to primarily provide a three dimensional map of the Milky Way. In 2023, Gaia published DR3 (\citealt{2023A&A...674A..26C}), including radius, mass, age, chemical abundances and more than 470 million stellar parameters (\teff, \logg~and \feh). For hot stars, ESP-HS determined rotation velocities by measuring line broadening on RVS spectra (\citealt{2023A&A...674A..28F}). 

We cross-matched LAMOST MRS \vsini~CATALOG and LAMOST LRS \vsini~CATALOG with Gaia DR3, respectively, and obtained 4526 common stars for MRS and 7013 common stars for LRS. Figure \ref{fig:mrs-gaia} shows the comparison between the MRS catalog and Gaia DR3 catalog, and Figure \ref{fig:lrs-gaia} shows the comparison of the LRS catalog to the Gaia DR3 catalog. The results in this work generally agree with Gaia \vsini~ with scatters of 12.2 \kms~and 24.7 \kms~for MRS and LRS, respectively. The scatter of MRS is relatively larger than that of APOGEE DR17 (3.10 \kms), because ESP-HS only provided \vsini~for hot stars, and \vsini-related information for hot stars in the wavelength of 8450 -- 8720 \AA \ is poor \citep{2023A&A...674A..28F} .

\subsubsection{Comparison with SUN}

\cite{2021ApJS..257...22S} used a machine learning algorithm to obtain stellar rotation, which trained a model with the theoretical kurucz spectra, and was used to calculate \vsini~of observed spectra.

We cross-matched LAMOST MRS \vsini~CATALOG and LAMOST LRS \vsini~CATALOG with the SUN catalog (\citealt{2021ApJS..257...22S}), and obtained 8267 and 2628 common stars, respectively. We compared the results in the two LAMOST catalogs with those in SUN, which are shown in Figure \ref{fig:mrs-sun1}, \ref{fig:lrs-sun}. The offset and scatter between LAMOST MRS and SUN are 2.2 \kms~and 6.5 \kms, and those of LRS are 1.2 \kms~and 28.4 \kms, which indicates the results are consistent with SUN's results, and the LRS has relatively large scatter.



\section{Summary} \label{sec:summary}
In this work, we used $\chi^2$ minimization to derive the best-matched \vsini~under the constraints of atmospheric parameters determined by LAMOST stellar parameter pipeline (LASP), through comparing the observed spectra to the PHOENIX synthetic spectra. We validated this method on APOGEE DR17, integrated it into LASP, and applied it to LAMOST MRS and LRS spectra. Eventually, we obtained \vsini~catalogs for both MRS and LRS:
\begin{enumerate}

  \item The \vsini~determination method. We used minimum $\chi^2$ algorithm to calculate \vsini~through comparing the observed spectrum with the reference spectra from PHOENIX grids based on the atmospheric parameters have already been determined by LASP. We firstly generated the synthetic reference spectrum with the theoretical PHOENIX grids and decreased the reference spectrum resolution to the observed spectrum resolution. Then, the reference spectrum was convolved by the broadening kernel with different \vsini~taken from \cite{Gray05}, and we finally obtained \vsini~by the minimum $\chi^2$ between the observed spectrum and those convolved reference spectra. 
  
  \item Method validation. We selected 27,000 APOGEE DR17 spectra with \teff~> 5000 K and \vsini~> 8 \kms~to calculate their \vsini~with the method proposed in this work, and the offset and scatter are 0.35 \kms~and 1.27 \kms~compared to APOGEE DR17. The resolution of APOGEE spectra were reduced to that of LAMOST MRS spectra (R $\sim$ 7500), \vsini~were calculated for these spectra, compared to APOGEE results, the offset and scatter are 3 \kms~and 8.79 \kms, respectively, which indicates that \vsini~ given by this work are consistent with APOGEE DR17 and the \vsini~determination method can be used for LAMOST spectra.

  \item Application to LAMOST. We integrated the method into LASP, and applied it to both MRS and LRS spectra based on the determined atmospherical parameters of LASP. As mentioned in \citep{2016A&A...594A..39F}, stars with \vsini~< 27 \kms~of MRS and \vsini~< 110 \kms~of LRS were removed, and we established two \vsini~catalogs, i.e., LAMOST MRS \vsini~CATALOG and LAMOST LRS \vsini~CATALOG, for 121,698 MRS stars and 80,108 LRS stars. For both MRS and LRS, the intrinsic precisions of \vsini~were obtained by the repeat observations, and they are $\sim$ 4.0 \kms~for MRS and $\sim$ 10.0 \kms~for LRS as S/N > 50.

  We cross-matched the two LAMOST \vsini~catalogs finding rotational velocities of 6000 common stars are consistent with an offset and scatter of 3.45 \kms~and 20.76 \kms~. We found that \vsini~differences of the two catalogs increase at \teff~> 8000 K, because the spectral lines become weaker in MRS blue arm spectra (4950 -- 5300 \AA\ ) when \teff~increases. LAMOST MRS \vsini~CATALOG were compared with APOGEE DR17, Gaia DR3 and SUN, respectively, \vsini~in MRS are consistent with the other three catalogs and the scatters are 3.1 \kms, 12.2 \kms~and 6.5 \kms, respectively. We also compared LAMOST LRS \vsini~CATALOG with Gaia DR3 and SUN, \vsini~finding scatters are 24.7 \kms~and 28.4 \kms. We noticed that the dispersion of the MRS catalog and Gaia DR3 is obviously larger than those between MRS and the other two catalogs, because Gaia only provided \vsini~for hot stars and the \vsini-related information of these stars are poor in the wavelength range of 8450 -- 8720 \AA\ \citep{2023A&A...674A..28F}.
  
  In addition, stars with -0.5 $\leq$\ \feh~$\leq$ 0.5 dex in the LAMOST MRS \vsini~CATALOG were selected and separated into dwarfs and giants, and we found that \vsini~of both dwarfs and giants decrease as \teff~drops, the former transition occurs at \teff~$\sim$ 7000 K, and the latter occurs at 6500 K, such a result is consistent with the rotation behavior presented in \cite{1967ApJ...150..551K} \& \cite{1989ApJ...347.1021G}.
  
  



\end{enumerate}

\begin{acknowledgements}
This work is supported by National Key R\&D Program of China No. 2019YFA0405502, the National Science Foundation of China (grant Nos. U1931209, 12273078 and 12073046), and China Manned Space Project (Nos. CMS-CSST-2021-A10). We thank Hou Wen, Wang Rui for helpful discussions. Guoshoujing Telescope (the Large Sky Area Multi-Object Fiber Spectroscopic Telescope, LAMOST) is a National Major Scientific Project built by the Chinese Academy of Sciences. Funding for the project has been provided by the National Development and Reform Commission. LAMOST is operated and managed by the National Astronomical Observatories, the Chinese Academy of Sciences. This research makes use of data from the European Space Agency (ESA) mission Gaia, processed by the Gaia Data Processing and Analysis Consortium.

\end{acknowledgements}




\begin{thebibliography}{}
\expandafter\ifx\csname natexlab\endcsname\relax\def\natexlab#1{#1}\fi
\providecommand{\url}[1]{\href{#1}{#1}}
\providecommand{\dodoi}[1]{doi:~\href{http://doi.org/#1}{\nolinkurl{#1}}}
\providecommand{\doeprint}[1]{\href{http://ascl.net/#1}{\nolinkurl{http://ascl.net/#1}}}
\providecommand{\doarXiv}[1]{\href{https://arxiv.org/abs/#1}{\nolinkurl{https://arxiv.org/abs/#1}}}

\end{thebibliography}


\begin{thebibliography}{}


\bibitem[Burkhart(1979)]{1979A&A....74...38B} Burkhart, C.\ 1979, \aap, 74, 38
\bibitem[Clementini et al.(2016)]{2016A&A...595A.133C} Clementini, G., Ripepi, V., Leccia, S., et al.\ 2016, \aap, 595, A133
\bibitem[Creevey et al.(2023)]{2023A&A...674A..26C} Creevey, O.~L., Sordo, R., Pailler, F., et al.\ 2023, \aap, 674, A26
\bibitem[Cui et al.(2012)]{2012RAA....12.1197C} Cui, X.-Q., Zhao, Y.-H., Chu, Y.-Q., et al.\ 2012, Research in Astronomy and Astrophysics, 12, 1197
\bibitem[Delfosse et al.(1998)]{1998A&A...331..581D} Delfosse, X., Forveille, T., Perrier, C., et al.\ 1998, \aap, 331, 581
\bibitem[Du et al.(2019)]{2019ApJS..240...10D} Du, B., Luo, A.-L., Zuo, F., et al.\ 2019, \apjs, 240, 10

\bibitem[Du et al.(2021)]{2021RAA....21..202D} Du, B., Luo, A.-L., Zhang, S., et al.\ 2021, Research in Astronomy and Astrophysics, 21, 202
\bibitem[El-Badry et al.(2021)]{2021MNRAS.506.2269E} El-Badry, K., Rix, H.-W., \& Heintz, T.~M.\ 2021, \mnras, 506, 2269
\bibitem[Fouesneau et al.(2023)]{2023A&A...674A..28F} Fouesneau, M., Fr{\'e}mat, Y., Andrae, R., et al.\ 2023, \aap, 674, A28
\bibitem[Frasca et al.(2016)]{2016A&A...594A..39F} Frasca, A., Molenda-{\.Z}akowicz, J., De Cat, P., et al.\ 2016, \aap, 594, A39
\bibitem[Fukuda(1982)]{1982PASP...94..271F} Fukuda, I.\ 1982, \pasp, 94, 271
\bibitem[Gaia Collaboration et al.(2016)]{2016A&A...595A...1G} Gaia Collaboration, Prusti, T., de Bruijne, J.~H.~J., et al.\ 2016, \aap, 595, A1
\bibitem[Gaia Collaboration et al.(2021)]{2021A&A...649A...1G} Gaia Collaboration, Brown, A.~G.~A., Vallenari, A., et al.\ 2021, \aap, 649, A1
\bibitem[Garc{\'\i}a P{\'e}rez et al.(2016)]{2016AJ....151..144G} Garc{\'\i}a P{\'e}rez, A.~E., Allende Prieto, C., Holtzman, J.~A., et al.\ 2016, \aj, 151, 144

\bibitem[Gray(1989)]{1989ApJ...347.1021G} Gray, D.~F.\ 1989, \apj, 347, 1021
\bibitem[Gray(2005)] {Gray05} Gray, D. F. 2005, The Observation and Analysis of Stellar Photospheres, 3rd edn. (Cambridge University Press)


\bibitem[Husser et al.(2013)]{2013A&A...553A...6H} Husser, T.-O., Wende-von Berg, S., Dreizler, S., et al.\ 2013, \aap, 553, A6

\bibitem[J{\"o}nsson et al.(2020)]{2020AJ....160..120J} J{\"o}nsson, H., Holtzman, J.~A., Allende Prieto, C., et al.\ 2020, \aj, 160, 120

\bibitem[Kamann et al.(2020)]{2020MNRAS.492.2177K} Kamann, S., Bastian, N., Gossage, S., et al.\ 2020, \mnras, 492, 2177
\bibitem[Kawaler(1988)]{1988ApJ...333..236K} Kawaler, S.~D.\ 1988, \apj, 333, 236
\bibitem[Kraft(1967)]{1967ApJ...150..551K} Kraft, R.~P.\ 1967, \apj, 150, 551
\bibitem[Kounkel et al.(2023)]{2023AJ....165..182K} Kounkel, M., Stassun, K.~G., Hillenbrand, L.~A., et al.\ 2023, \aj, 165, 182

\bibitem[Levenhagen(2014)]{2014ApJ...797...29L} Levenhagen, R.~S.\ 2014, \apj, 797, 29

\bibitem[Luo et al.(2015)]{2015RAA....15.1095L} Luo, A.-L., Zhao, Y.-H., Zhao, G., et al.\ 2015, Research in Astronomy and Astrophysics, 15, 1095
\bibitem[Lu et al.(2023)]{2023arXiv231014990L} Lu, Y., Angus, R., Foreman-Mackey, D., Hattori, S., 2023arXiv231014990L
\bibitem[Melo et al.(2001)]{2001A&A...375..851M} Melo, C.~H.~F., Pasquini, L., \& De Medeiros, J.~R.\ 2001, \aap, 375, 851

\bibitem[Nofi et al.(2021)]{2021ApJ...911..138N} Nofi, L.~A., Johns-Krull, C.~M., L{\'o}pez-Valdivia, R., et al.\ 2021, \apj, 911, 138
\bibitem[Penoyre et al.(2022)]{2022MNRAS.513.5270P} Penoyre, Z., Belokurov, V., \& Evans, N.~W.\ 2022, \mnras, 513, 5270
\bibitem[Recio-Blanco et al.(2016)]{2016A&A...585A..93R} Recio-Blanco, A., de Laverny, P., Allende Prieto, C., et al.\ 2016, \aap, 585, A93
\bibitem[Reiners \& Basri(2008)]{2008ApJ...684.1390R} Reiners, A. \& Basri, G.\ 2008, \apj, 684, 1390
\bibitem[Royer et al.(2002)]{2002A&A...393..897R} Royer, F., Grenier, S., Baylac, M.-O., et al.\ 2002, \aap, 393, 897
\bibitem[Schatzman(1962)]{1962AnAp...25...18S} Schatzman, E.\ 1962, Annales d'Astrophysique, 25, 18
\bibitem[Shang et al.(2022)]{2022ApJS..259...63S} Shang, L.-H., Luo, A.-L., Wang, L., et al.\ 2022, \apjs, 259, 63
\bibitem[Slettebak(1949)]{1949ApJ...110..498S} Slettebak, A.\ 1949, \apj, 110, 498
\bibitem[Slettebak(1954)]{1954ApJ...119..146S} Slettebak, A.\ 1954, \apj, 119, 146

\bibitem[Slettebak(1955)]{1955ApJ...121..653S} Slettebak, A.\ 1955, \apj, 121, 653
\bibitem[Slettebak et al.(1975)]{1975ApJS...29..137S} Slettebak, A., Collins, G.~W., Boyce, P.~B., et al.\ 1975, \apjs, 29, 137

\bibitem[Smith \& Gray(1976)]{1976PASP...88..809S} Smith, M.~A. \& Gray, D.~F.\ 1976, \pasp, 88, 809
\bibitem[Su \& Cui(2004)]{2004ChJAA...4....1S} Su, D.-Q. \& Cui, X.-Q.\ 2004, \cjaa, 4, 1
\bibitem[Sun et al.(2019)]{2019ApJ...876..113S} Sun, W., de Grijs, R., Deng, L., et al.\ 2019, \apj, 876, 113
\bibitem[Sun et al.(2021)]{2021ApJS..257...22S} Sun, W., Duan, X.-W., Deng, L., et al.\ 2021, \apjs, 257, 22

\bibitem[Takeda(2020)]{2020PASJ...72...10T} Takeda, Y.\ 2020, \pasj, 72, 10
\bibitem[Wang et al.(1996)]{1996ApOpt..35.5155W} Wang, S.-G., Su, D.-Q., Chu, Y.-Q., et al.\ 1996, \ao, 35, 5155
\bibitem[Wang et al.(2019)]{2019ApJS..244...27W} Wang, R., Luo, A.-L., Chen, J.-J., et al.\ 2019, \apjs, 244, 27
\bibitem[Whiting et al.(2023)]{2023AJ....165..193W} Whiting, M.~L., Hill, J.~B., Bromley, B.~C., et al.\ 2023, \aj, 165, 193

\bibitem[Zhang et al.(2022)]{2022ApJS..259...38Z} Zhang, Y.-J., Hou, W., Luo, A.-L., et al.\ 2022, \apjs, 259, 38
\bibitem[Zhao \& Newberg(2006)]{2006astro.ph.12034Z} Zhao, C. \& Newberg, H.~J.\ 2006, astro-ph/0612034
\bibitem[Zhao et al.(2012)]{2012RAA....12..723Z} Zhao, G., Zhao, Y.-H., Chu, Y.-Q., et al.\ 2012, Research in Astronomy and Astrophysics, 12, 723

\end{thebibliography}
\end{document}